\newcommand{\PaperTitle}{Waiting at the front door: Continuous monitoring of latency in the host network stack}
\documentclass[sigconf,letterpaper,nonacm,authorversion]{acmart}

\usepackage[T1]{fontenc}
\usepackage[utf8]{inputenc}
\usepackage{csquotes}
\usepackage[english]{babel}
\usepackage{microtype}
\usepackage{multirow}
\usepackage{subcaption}
\usepackage{siunitx}
\usepackage{fancyhdr}

\AtBeginDocument{%
  }

\copyrightyear{2026}
\setcopyright{cc}
\setcctype{by}

\begin{document}
% \fancyfoot[C]{\\ \small This is the author's version of the work. It is posted here for your personal use. Not for redistribution.}

\title{\PaperTitle}


\author{Simon Sundberg}
\email{simon.sundberg@kau.se}
\orcid{0000-0002-3570-9525}
\affiliation{%
  \institution{Karlstad University}
  \city{Karlstad}
  %\state{}
  \country{Sweden}
}

\author{Anna Brunstrom}
\email{anna.brunstrom@kau.se}
\orcid{0000-0001-7311-9334}
\affiliation{%
  \institution{Karlstad University}
  \city{Karlstad}
  %\state{}
  \country{Sweden}
}

\author{Simone Ferlin-Reiter}
\email{sferlin@rehat.com}
%\email{simon.ferlin@kau.se}
\orcid{0000-0002-0722-2656}
\affiliation{%
  \institution{Red Hat}
  \city{Copenhagen}
  \country{Denmark}
}
\affiliation{%
  \institution{Karlstad University}
  \city{Karlstad}
  %\state{}
  \country{Sweden}
}

\author{Jesper Dangaard Brouer}
\email{jesper@cloudflare.com}
\orcid{0009-0001-1467-0684}
\affiliation{%
  \institution{Cloudflare}
  \city{Copenhagen}
  \country{Denmark}
}

\author{Toke Høiland-Jørgensen}
\email{toke@redhat.com}
\orcid{0000-0001-5241-6815}
\affiliation{%
  \institution{Red Hat}
  \city{Copenhagen}
  \country{Denmark}
}

%\renewcommand{\shortauthors}{Sundberg et al.}

\begin{abstract}
With networking moving into the sub-millisecond latency domain, latency in the end host itself can become a significant barrier to achieving consistently low application latency. Both the physical interconnect between the network card and the CPU, the kernel network stack, and the scheduling of applications themselves can be considerable sources of latency. Previous work has studied host latency at various levels, yet there remains a lack of methods and tools to continuously monitor host latency in production. To remedy this, we present netstacklat, a monitoring tool that captures latency at several points in the host network, from the early parts of the Linux kernel network stack all the way until the application reads the data. We evaluate netstacklat in a testbed, demonstrating its ability to capture host latency across 144 variations of HTTP workloads for Nginx and Apache, while also showing how the low monitoring overhead does not inflate tail latency by more than 6\%, where previous monitoring solutions increase it by over 100\%. Furthermore, we share our initial findings from deploying netstacklat in Cloudflare's global CDN network.
\end{abstract}




\maketitle

\section{Introduction}
Traditionally, end-to-end latency has been attributed mainly to network congestion, where queueing occurs primarily in the network fabric. However, as network speeds continue to outpace host I/O and memory bandwidth improvements, hosts themselves have become critical points of contention. Many studies have shown that the host can have a significant impact on the end-to-end latency, for example due to congestion in the interconnect between the Network Interface Card (NIC) and CPU~\cite{agarwalUnderstandingHostInterconnect2022, agarwalHostCongestionControl2023, vuppalapatiUnderstandingHostNetwork2024}, architectural inefficiencies in the Linux network stack~\cite{liTalesTailHardware2014, caiUnderstandingHostNetwork2021, caiUsTailLatency2022, awamotoOpeningKernelBypassTCP2025, jasnyWakeUpCallKernelBypass2025}, interrupt processing~\cite{beifussStudyNetworkingSoftware2015, gallenmullerDuckedTailsTrimming2021, caiKernelVsUserlevel2023}, and container and virtualization overhead~\cite{emmerichStudyNetworkStack2014, oljiraAnalysisNetworkLatency2016, qiSPRIGHTExtractingServer2022, daichendtApplicabilityHardwaresupportedContainers2024}. As such, the host network latency, which encompasses the delay from a packet arriving at the NIC, being transferred to the CPU, traversing the kernel's network stack, and finally being read by the application, is becoming increasingly important as modern networks push for sub-millisecond performance. Being able to continuously monitor the host network latency would therefore be highly useful to identify local latency issues that could have a large impact on the end-to-end performance.

Yet, despite this wealth of work examining the impact the host has on network latency, there remains a lack of methods to continuously monitor host network latency. While actively measuring end-to-end application latency~\cite{emmerichStudyNetworkStack2014, beifussStudyNetworkingSoftware2015, oljiraAnalysisNetworkLatency2016, caiUsTailLatency2022, qiSPRIGHTExtractingServer2022, caiKernelVsUserlevel2023, awamotoOpeningKernelBypassTCP2025, jasnyWakeUpCallKernelBypass2025}, analyzing packet captures from external tap devices~\cite{gallenmullerDuckedTailsTrimming2021, daichendtApplicabilityHardwaresupportedContainers2024}, or patching the kernel to log packet timestamps~\cite{liTalesTailHardware2014, caiUnderstandingHostNetwork2021} are suitable methods to infer host latency in controlled experiments, they are of little use for continuously monitoring the host network latency on servers running in production. To address this gap, we therefore make the following contributions:

\begin{enumerate}
    \item We design and implement netstacklat, a tool able to monitor the latency buildup as packets traverse the ingress network stack, from early parts in the kernel's receive path all the way up to application reception. By combining the Linux kernel's timestamping mechanism with eBPF, netstacklat is able to track host latency for every packet system-wide with a low enough footprint to enable continuous monitoring. 
    % Netstacklat can either be used by itself for ad-hoc monitoring directly from the terminal, or combined with ebpf-exporter to provide Prometheus metrics.
    \item We evaluate netstacklat in a testbed across 144 Nginx and Apache workloads, showing how netstacklat can track long tail latencies at several levels in the kernel network stack. Furthermore, we demonstrate that netstacklat has reasonably low overhead, only averaging 0.81\% CPU utilization across all workloads, and only inflating end-to-end tail latency by 6\% in a heavily loaded scenario where other state-of-the-art solutions increase tail latency by over 100\%.
    \item We present early insights from deploying netstacklat at Cloudflare, illustrating how host network latency increases as a server experiences progressively higher load. We also show how netstacklat uncovered an anomalous event where the host latency continuously increased over a 3-hour period before returning to normal.
\end{enumerate}

The rest of the paper is structured as follows: Section~\ref{sec:related-work} covers existing work on monitoring host network latency. Section~\ref{sec:background} provides a brief background on the Linux ingress network stack and its support for packet timestamping. Section~\ref{sec:netstacklat-design} describes the design of netstacklat. In Section~\ref{sec:workload-comparison}, we evaluate netstacklat on a testbed with several HTTP workloads and show how the network stack latency and the monitoring overhead vary across all configurations. Section~\ref{sec:tool-comparison} compares the monitoring overhead of netstacklat with other tools that can be used for inferring network stack latency. In Section~\ref{sec:cdn-deployment}, we share some initial insights from the deployment of netstacklat at Cloudflare. Section~\ref{sec:limitations} discusses current limitations with netstacklat and how future work can address them. Finally, Section~\ref{sec:conclusion} concludes the work.

\section{Related work}
\label{sec:related-work}
% Potential TODO: Put together a table comparing the tools as Simone suggested
 
% \item Fathom
While many works examine the impact of host network latency on the end-to-end performance, as covered in the introduction, the body of work focusing on monitoring aspects of host network latency is comparatively smaller. The most comprehensive monitoring solution in this vein is likely Google's Fathom~\cite{qureshiFathomUnderstandingDatacenter2023}. Fathom expands on Google's earlier work with Dapper~\cite{sigelmanDapperLargeScaleDistributed2010} for distributed monitoring of Remote Procedure Calls (RPCs) by collecting timestamps at several places in the network stack and application layer on both the client and server side. These timestamp-enriched traces provide a detailed breakdown of latency across the entire RPC chain. Furthermore, Fathom also offers automated root cause analysis for performance issues using a waterfall heuristic, and characterization of applications' typical network performance through a Gaussian Mixture Model (GMM). While Fathom offers extensive monitoring capabilities, it is limited to monitoring RPC calls using one of their instrumented RPC libraries in scenarios where both the client and server can be monitored. Furthermore, for the highly granular data collection to scale for fleet-wide monitoring, Fathom uses aggressive sampling, monitoring only 1:1000 or 1:128,000 RPC calls.
    
%\item NSight
A more generic tool for monitoring latency in the host network is NSight~\cite{haeckiHowDiagnoseNanosecond2022}, which utilizes the Intel-PT~\cite{intelcorporationIntel64IA322025} hardware profiler and NIC hardware timestamps to construct highly accurate per-message lifetimes with virtually no overhead. Unfortunately, at the time of writing, the NSight tool does not appear to be publicly available. While netstacklat cannot match neither the granularity of measurements nor the incredibly low overhead of Nsight, netstacklat offers an entirely software-based alternative that can be deployed without any kernel modifications. Furthermore, NSight is not well-suited for continuous monitoring due to limitations with the buffering in Intel-PT, and also produces upwards of 1 GB/s of monitoring data for a single host.

%\item lattrace
A closely related work to netstacklat is lattrace~\cite{maurerInvestigatingCausesJitter2021}, which has been a strong inspiration during netstacklat's development. Lattrace uses eBPF to record the time when SKBs reach many of the same points in the network stack as netstacklat. However, instead of aggregating latencies in-kernel, lattrace pushes timestamp records to a user space agent that then creates per-packet traces. While this provides more granular data than netstacklat's aggregated latency distributions, it requires frequent kernel-to-user space communication, resulting in significantly higher overhead. Furthermore, the large amount of output data makes lattrace unsuitable for long-term continuous monitoring. The recently presented Netto~\cite{miolaMeasuringCostLinux2024} also uses eBPF to monitor the kernel's network stack. However, Netto aims to monitor the CPU load for the network stack rather than the per-packet latency. %, with the authors finding that sampling the kernel call stack is the most suitable approach for their goal.

% Nameless tool from "Where does the time go? Characterizing tail latency in memcached"
In~\cite{blakeWhereDoesTime2015}, the authors implement a custom tool to measure the tail latency for Memcached, including latency components within the kernel network stack. This tool measures the cumulative latency between the early parts of the network stack and several later points using a timestamp kept as part of the packet metadata in the kernel, and then aggregates the latencies into distributions (although the paper does not provide any details on the aggregation mechanism). While this design is similar to that of netstacklat, netstacklat differs by using eBPF instead of SystemTap to instrument the kernel, and by reusing the kernel's existing time stamping mechanism rather than writing its own timestamps to the packet metadata structure. These differences make netstacklat applicable to a wider range of systems, because it does not require modifying the kernel to support SystemTap, and avoids interfering with other timestamping uses on the system.  

%\item Kyanos
Other relevant tools for monitoring host network stack latency have also appeared outside of academia. Kyanos~\cite{kyanos-webpage} uses eBPF to capture timestamps at several points in the network stack to create latency traces. Kyanos is application protocol aware, creating latency traces for request-response pairs for HTTP 1.1, MySQL, and Redis. This results in a more complete view of the end-to-end latency breakdown than what netstacklat provides, although it also limits kyanos to only monitor applications using one of the three supported protocols. Furthermore, like the aforementioned lattrace, kyanos pushes individual timestamp records to user space for processing and therefore also has much higher overhead than netstacklat. 

Additionally, tools such as pwru~\cite{liangPwruLinuxKernel2024} and retis~\cite{tenartChallengesLimitationsDebugging2025} also utilize eBPF to trace packets across the network stack. 
% Both pwru and retis can collect a vast amount of information about the packet state with additional context from all kernel functions processing SKBs. 
However, these tools are designed for debugging the network stack rather than continuous monitoring, collecting vast amounts of highly detailed data when overhead is of little concern. We consider netstacklat to be a complement to these network debugging tools. Netstacklat can be used to detect when latency builds up in the local host network and provides a rough indication of where the issue arises. Tools such as pwru and retis can then be used to further troubleshoot and identify the root cause. We compare the overhead of netstacklat, lattrace, kaynos, retis, and pwru in Section~\ref{sec:tool-comparison}.

\section{Background}
\label{sec:background}
This section gives a brief introduction to two core aspects of the Linux network stack relevant for understanding how netstacklat works: The packet processing path (Section~\ref{sec:linux:kernel:network:stack}) and the kernel's packet timestamping facility (Section~\ref{sec:linux:packet:timestamping}).

\subsection{Linux kernel network stack} 
\label{sec:linux:kernel:network:stack}

%The Linux kernel network stack has become a high-performance, modular subsystem with a well-defined layered structure: From device drivers and (new API) NAPI-based packet ingress to routing, netfilter, transport protocols, and the socket interface. Although Linux still maintains a general-purpose architecture, many optimizations introduced in recent years, such as Receive Packet Steering (RPS), Receive Flow Steering (RFS), eXpress Data Path (XDP), and eBPF, reflect similar principles of minimizing contention and enabling more direct, programmable packet-processing. As examples, frameworks like netfilter, eBPF, and XDP enable safe in-kernel programmability and (just in time) JIT-compiled datapaths, while remaining inside the kernel.

\begin{figure}
    \centering
    \includegraphics[width=\columnwidth]{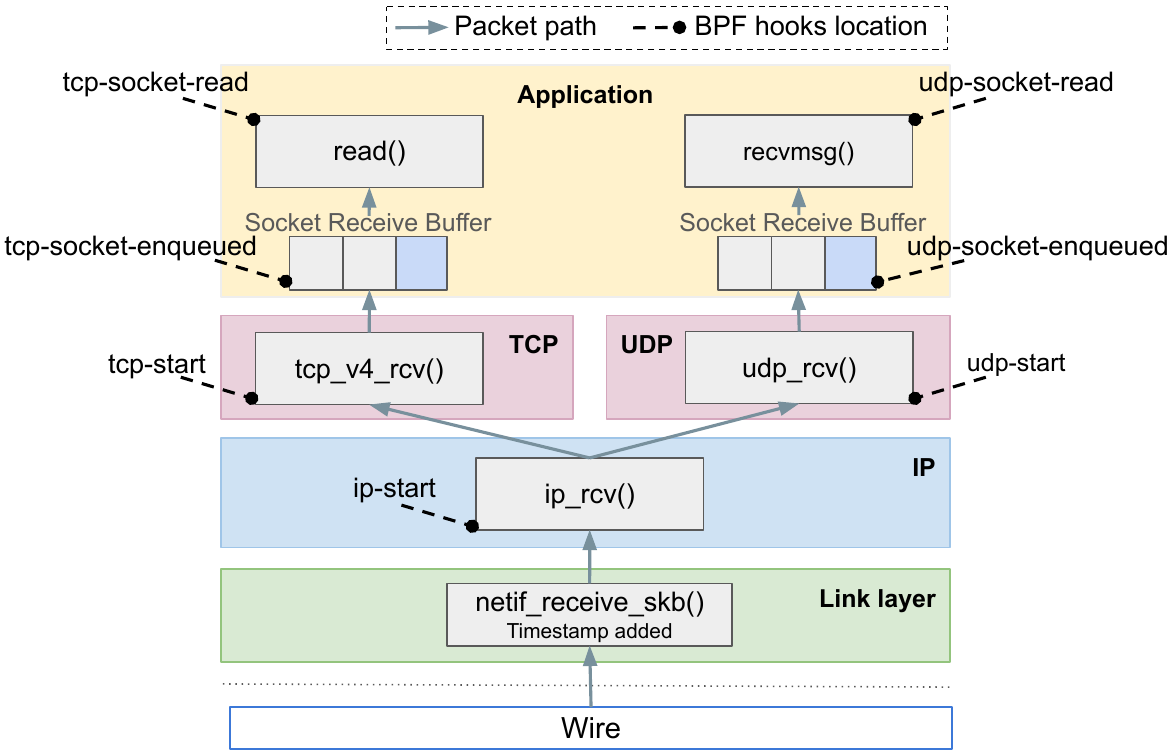}
    \Description{A schematic of the Linux ingress network stack, separated into the link, IP, transport, and application layers. Also shows how netstacklat probes functions at the start or the end of each layer.}
    \caption{Linux kernel networking stack ingress path with netstacklat probe points.}
    \label{fig:tcp:ingress:path}
\end{figure}
%Packets enter the stack from the wire at the bottom, and proceed through the IP protocol stack, and either TCP or UDP towards the application.  

The Linux kernel network stack represents incoming packets using a data structure called (for historical reasons) a \textit{socket buffer}, or \textit{SKB}. An SKB contains packet metadata such as protocol headers and checksum state, as well as references to the memory segments containing the actual packet data. When a network device receives a data frame, the driver either constructs a new SKB to represent the packet or attaches the frame data to an existing SKB, aggregating multiple data frames to a single logical packet. The driver then submits the SKB to the kernel's network stack for processing.

The path the packet takes through the core networking stack is illustrated in Figure~\ref{fig:tcp:ingress:path}, which also shows at which points netstacklat measures the network stack latency. The \texttt{netif\_receive\_skb()} function is the entry point called by the driver code. This function adds a timestamp to the packet (see the next section) and parses the packet headers to determine the path the packet will take through the networking stack. 

The Linux networking stack is modular and supports a wide variety of protocols, with each layer parsing the next bit of the packet header to extract metadata and determine the next processing layer. Figure~\ref{fig:tcp:ingress:path} shows the most common paths, comprised of the IP and TCP or UDP layers. For both TCP and UDP, the packet is ultimately enqueued into the socket's receive buffer, where it is kept until the application reads the data.

Reading the data from the socket is done through the appropriate system calls (such as \texttt{read()} or \texttt{recvmsg()}). These calls cause the kernel to dequeue one or more SKBs from the socket's receive queue, and to copy the payload to the data buffer provided by user space (or simply copy a descriptor, if the application has enabled \textit{zero-copy} operation). The kernel then updates the state of the socket (e.g., for a TCP connection the window and acknowledgments state), and finally releases or recycles the SKB.

\subsection{Linux packet timestamping} 
\label{sec:linux:packet:timestamping}
The Linux kernel provides several packet timestamping capabilities that applications can enable through the \texttt{SO\_TIME\-STAMPING}~\cite{thekerneldevelopmentcommunityTimestampingLinuxKernel2025} socket option. With these timestamping options, applications can retrieve timestamps attached to the socket buffers (SKBs) that represent packets in the kernel, or write timestamps before transmitting. One of the possible values of this socket option, \texttt{SOF\_TIME\-STAMPING\_RX\_SOFT\-WARE}, causes the kernel to generate a timestamp early in the network receive path for each SKB (as shown in the previous subsection). This timestamp can then be retrieved by the application along with the data that it reads from its socket, providing the application with a measure of when the network packet containing the read data first arrived at the host.

When the \texttt{SOF\_TIME\-STAMPING\_RX\_SOFTWARE} option is enabled on any socket, the kernel timestamps all SKBs it processes, since at the early processing stage where the timestamp is added, the kernel does not yet know which socket (if any) the SKB will be delivered to. The timestamp is stored in the SKB structure itself (in the \texttt{tstamp} field), which means it will travel with the packet throughout the networking stack. Netstacklat takes advantage of this to track the packet timings, as detailed in the following section.

\section{Netstacklat design}
\label{sec:netstacklat-design}
Netstacklat aims to enable efficient monitoring of how latency builds up from the start of the kernel network stack up to the application reception of data. For the remainder of this paper, we will refer to this latency as network stack latency. We achieve this by collecting the latency distribution at several layers in the network stack. Netstacklat can either be used as a standalone tool in the terminal for ad-hoc monitoring sessions, or set up together with ebpf-exporter~\cite{babrouEbpf_exporter2025} to export data to Prometheus for continuous, always-on monitoring. We have made the source code for netstacklat publicly available\footnote{\url{https://github.com/xdp-project/bpf-examples/tree/main/netstacklat}}. 

In the remainder of this section, we explain the design of netstacklat in detail. While the core idea (Section~\ref{sec:netstacklat-core-idea}) is relatively simple, we encountered several challenges while turning this idea into a tool that is usable across a wide range of production systems. These challenges are treated in each of the following subsections and include: the selection of suitable points to probe in the network stack (Section~\ref{sec:netstacklat-hooks}); efficiently aggregating the latency values to produce a manageable amount of monitoring output (Section~\ref{sec:netstacklat-aggregation}); and avoiding reporting inflated network stack latency from head-of-line blocking caused by the external network (Section~\ref{sec:netstacklat-tcp-hol-filtering}).

\subsection{Core idea}
\label{sec:netstacklat-core-idea}
The core idea behind netstacklat is simple: By observing SKBs at several strategically selected points across the kernel network stack, we can break down how the network stack latency builds up as the SKBs traverse the stack.

To achieve this, we use eBPF~\cite{gbadamosiEBPFRuntimeLinux2024} to attach small programs to key functions in the kernel network stack. Specifically, we use eBPF tracing programs (\texttt{BPF\_PROG\_TY\-PE\_TRACING}) with the \texttt{fentry} or \texttt{fexit} attach points. Thanks to a trampoline mechanism~\cite{reimerinkTrampolinesEBPFDocs2025}, the kernel can, with very small overhead, jump to a \texttt{fentry} or \texttt{fexit} program at the start or end of the kernel function that the probe is attached to. Furthermore, \texttt{fentry} and \texttt{fexit} programs have direct memory access to all arguments provided to the probed function. This means that by attaching \texttt{fentry} and \texttt{fexit} programs to kernel functions that process SKBs, netstacklat is able to efficiently observe the packets as they traverse the stack. Figure~\ref{fig:tcp:ingress:path} shows a high-level view of the Linux network stack ingress path and the points where netstacklat attaches its probes.

While prior solutions~\cite{maurerInvestigatingCausesJitter2021, kyanos-webpage, liangPwruLinuxKernel2024, tenartChallengesLimitationsDebugging2025} also use eBPF to observe SKBs across the kernel, we use a novel approach for determining the network stack latency at each probe point. In prior solutions, the eBPF programs record the time at each observation point and push those timestamp records to user space. A user space program then matches all timestamp records for each SKB and calculates the delay between the timestamps. In contrast, we instead utilize the SKB receive timestamp generated by \texttt{SOF\_TIMESTAMPING\_RX\_SOFTWARE} (Section~\ref{sec:linux:packet:timestamping}). With the SKB timestamp, the network stack latency can at any point be calculated as the difference between the current time and the SKB timestamp. This allows netstacklat to directly calculate the latency in the eBPF programs, which is key for enabling in-kernel aggregation (Section~\ref{sec:netstacklat-aggregation}). Furthermore, by reusing the kernel's SKB timestamp, we avoid the need to keep any external state for each SKB and the associated overhead from looking up and managing that state at each probe point. 

As all latencies are calculated relative to the SKB receive timestamp, netstacklat reports the cumulative latency up to each probe point rather than the difference from the previous probe. However, by comparing the aggregated values for each point, it is still possible to infer where in the stack the latency accumulates, as detailed in the following section.

\subsection{The netstacklat probe points}
\label{sec:netstacklat-hooks}
There are several considerations to take into account when selecting which kernel functions in the network stack we should probe. The probed functions should have access to the SKB timestamp, be stable across kernel versions, and correspond to clear logical boundaries to make the results easy to interpret. To select the probe points used by netstacklat, we took some inspiration from previous works, like kyanos and, in particular, lattrace, while making some adjustments due to the cumulative nature of the measurements performed by netstacklat.
% All probed functions have been part of the Linux kernel since at least version 4.19 %, and we therefore deem it likely that they will remain in the kernel for the foreseeable future.
This selection has resulted in netstacklat recording the network stack latency at four different layers, as shown in Figure~\ref{fig:tcp:ingress:path}, with the probed functions listed in Table~\ref{tab:hooks}:

% Potential TODO: Comment on that probes should run for most SKBs, regardless of their path through the network stack, to avoid blindspots. Meanwhile, the probes should not run multiple times for the same SKB, i.e. the kernel functions mapped to the same probe name should not overlap.

\begin{table}
    \centering
    \caption{Netstacklat probe points}
    \label{tab:hooks}
    \begin{tabular}{l l}
        \toprule
        Netstacklat label   &  Kernel function\\
        \midrule
        ip-start            & ip\_rcv\_core, ip6\_rcv\_core   \\ % since kernel 4.19
        %\midrule
        \addlinespace
        udp-start           & udp\_rcv, udpv6\_rcv            \\ % since kernel 0.98/2.1
        tcp-start           & tcp\_v4\_rcv, tcp\_v6\_rcv      \\ % since kernel 2.1
        %\midrule
        \addlinespace
        udp-socket-enqueued & \_\_udp\_enqueue\_schedule\_skb \\ % since kernel 4.10
        tcp-socket-enqueued & tcp\_queue\_rcv                 \\ % since kernel 3.5
        %\midrule
        \addlinespace
        udp-socket-read     & skb\_consume\_udp               \\ % since kernel 4.10
        tcp-socket-read     & tcp\_recv\_timestamp            \\ % since kernel 4.14
        \bottomrule
    \end{tabular}
\end{table}

\begin{description}
\item[ip-start]
The point at which the SKB has reached the start of IP layer processing. Latency at this point comes from early parts of the network stack processing, such as VLAN and traffic control (tc) handling.

%\item[\{udp,tcp\}-start]
\item[udp-start and tcp-start]
The SKB has reached the start of UDP or TCP processing. Latency at this point includes IP-layer processing, such as routing and the Netfilter firewall subsystem, in addition to the latency from ip-start.

\item[socket-enqueued]
The SKB has now been enqueued to the application UDP or TCP socket. This is the end of network stack processing, and the SKB payload is now ready to be read by the application. Latency here includes the UDP or TCP layer processing, as well as the latency from udp-start and tcp-start.  

\item[socket-read]
The SKB payload has now been read from the UDP or TCP socket into the application user space buffer. Unlike the previous layers, this runs in response to the application requesting to read data from the socket rather than the arrival of a network packet. 
%The latency here effectively includes the total network stack latency before the application can start processing received data and may include...
In addition to the network stack latency from the socket-enqueued probes, the latency here also includes any delays from CPU scheduling.
Furthermore, the latency may also be inflated by the application itself if it performs or waits on any tasks before attempting to read from the socket.
\end{description}

By capturing network stack latency at these four layers, netstacklat can break down where in the network stack significant latency increases occur. As each probe point reports the cumulative latency up to that point, this breakdown should be performed top-down. If, for example, tcp-socket-read latency is high and tcp-socket-enqueued is low, it indicates an issue at the application layer. Meanwhile, if tcp-socket-enqueued and tcp-start are also high while ip-start is low, it indicates that the issue is within the IP layer.

% The way we interpret netstacklat's latency measurements, taking an example of TCP's ingress path, is the following: If the latency observed at \texttt{tcp-socket-read} is high, we know that this is happening somewhere before that in the kernel. If \texttt{tcp-socket-enqueued} latency is at the same time low, then the latency occurs between these two points, meaning that it is due to an application that is taking a long time to read the data already delivered to the socket. This points to two possibilities: It could be either an issue with the application itself, or simply that the application is waiting on being scheduled to run on the CPU. If, on the other hand, both \texttt{tcp-socket-read} and \texttt{tcp-socket-enqueued} show high latency, latency must be coming from somewhere earlier in the path. If the latency issue is between \texttt{tcp-socket-enqueued} and \texttt{tcp-start}, the TCP layer processing is slow, and if it is between \texttt{ip-start} and \texttt{tcp-start}, the latency is added at the IP-layer, for example by the Netfilter firewall subsystem. This way, we can compare latency readings from the different hook points with netstacklat, which allows us to pinpoint where in the stack the latency is added. 

As each probe delays the progress of the SKB through the network stack, the number of probes is a tradeoff between the granularity of the latency breakdown and the additional latency and overhead added to the network stack processing. In Section~\ref{sec:tool-comparison}, we show that on our testbed netstacklat adds around 225ns per probe, so the 4 probe points can be expected to add under 1us of additional processing time for an SKB. However, as each netstacklat probe point operates independently, any probe point can easily be disabled if the latency at the corresponding level is not of interest.

\subsection{Aggregation of latency values}
\label{sec:netstacklat-aggregation}
One key design choice to make netstacklat suitable for continuous monitoring is to directly aggregate the latency measurements in-kernel rather than reporting separate latency samples. This has two key benefits. Firstly, aggregation eliminates the need for eBPF programs to report each individual value to a user space agent.
%, which is typically done via a ring buffer, e.g \texttt{BPF\_MAP\_TYPE\_RINGBUF} or \texttt{BPF\_MAP\_TYPE\_PERF\_EVENT\_ARRAY}. 
This not only eliminates frequent communication between kernel space and user space, but also completely decouples the recording of values in the eBPF programs from the fetching of the values by the user space agent. Therefore, an increase in event frequency does not need to incur any increase in processing latency for the user space agent, only the eBPF programs. 
We note that netstacklat's mechanism for calculating the latency in the eBPF programs (Section~\ref{sec:netstacklat-core-idea}) is crucial for enabling in-kernel aggregation. Prior solutions~\cite{maurerInvestigatingCausesJitter2021, kyanos-webpage, liangPwruLinuxKernel2024, tenartChallengesLimitationsDebugging2025} that instead record raw timestamps in the eBPF programs cannot meaningfully aggregate the data until after it has been processed in user space. %Prior solutions that only record timestamps in the eBPF programs cannot benefit from in-kernel aggregation, as the latency is calculated in user space once all timestamps have already been communicated. 
Secondly, aggregation keeps the size of the monitoring data manageable. Collecting timestamps or latency values together with associated metadata for every network packet would result in the monitoring data scaling proportionally to the packet rate. 
% Even using just 8 bytes per packet (e.g. a single 64-bit timestamp) could lead to hundreds of GB of data every day for a single busy server.
A suitable aggregation scheme can largely decouple the size of the monitoring data from the number of measurements it is based on.

For netstacklat, we chose to aggregate the data in histograms. This allows a wide range of distribution metrics, such as arbitrary quantiles, to be estimated with known error bounds. Histograms are also well supported in many popular monitoring platforms, simplifying the integration of netstacklat's monitoring data into existing monitoring pipelines. Using the flexible ebpf-exporter, we make it possible to export netstacklat's data to Prometheus without needing a separate custom exporter for netstacklat. Furthermore, histograms have several properties that make them well-suited for eBPF programs. Their fixed size avoids the need to dynamically allocate memory, and recording a value is very efficient as it only requires incrementing a counter for the corresponding histogram bin. As histograms are also mergeable, we can keep a separate copy of a histogram for each CPU core, allowing independent updates without costly synchronization between CPUs. The user space agent can then sum up the per-core histograms to get the complete distribution.

To support a wide range of potential latency values, from tens of nanoseconds to full seconds,
%, from tens of nanoseconds for the earlier parts of the network stack to in worst case seconds of latency until data is read by the application, we use an exponential histogram scheme. Specifically, we use a power-of-two histogram, where each bin is twice as wide as the previous one.
we use exponential power-of-two histograms, where each histogram bin is twice as wide as the previous one. 
This histogram scheme only provides coarse granularity with a maximum relative error of 
$\pm33.3\%$ from the bin midpoint. However, it allows us to cover the entire range of latency values with relatively few bins, which is important to limit the cardinality of the Prometheus time series.
By default, we use 35 histogram bins, covering the range from 1ns to 17.2s. A 36th bin stores the sum of all values, as typically provided for Prometheus histograms, allowing accurate calculation of the mean latency.

If ebpf-exporter is extended to support Prometheus's experimental native histograms~\cite{prometheusauthorsNativeHistogramsPrometheus2025}, higher histogram resolutions could be used. For instance, an exponential base of $2^{1/8}$ would limit the maximum relative error to $\pm 4.33$\% while covering the entire 1ns to 10s range with only 267 bins. However, the current base-2 resolution is still sufficient for netstacklat to detect the often very long tail latencies that can arise in the host network. As we show in Section~\ref{sec:workload-latency}, host latency can be inflated by several orders of magnitude under load.

While we opt for exponential histograms, the choice of aggregation scheme is not tightly coupled to the core idea of netstacklat and could be changed to fit different requirements. In theory, any fixed-space, mergeable data structure that allows independent updates across CPU cores could be used. For instance, t-digests~\cite{dunningComputingExtremelyAccurate2019} may offer more accurate estimations of high quantiles for a similar space tradeoff. However, t-digests are not supported in Prometheus and are more challenging to implement in eBPF, where floating-point calculations are not available. Furthermore, the error bounds for t-digests are lost once merged, which, for some very long-tailed distributions, can result in relative errors greater than 100\%~\cite{hartmannCircllhistLoglinearHistogram2020}.

\subsection{Handling TCP head-of-line blocking}
\label{sec:netstacklat-tcp-hol-filtering}
In general, the host should deliver arriving packets to the receiving application as soon as possible to minimize latency. However, for TCP traffic, all TCP segments must be delivered in order, and therefore, the host network stack may intentionally withhold TCP segments from the application until the missing segment gaps can be filled. The latency inflation caused by this TCP Head-Of-Line (HoL) blocking is primarily dependent on external network factors, e.g. the latency and packet loss across the network path to the other end host. Including the delay from this HoL blocking, as done by prior tools~\cite{maurerInvestigatingCausesJitter2021, kyanos-webpage, liangPwruLinuxKernel2024, tenartChallengesLimitationsDebugging2025}, can therefore be misleading when focusing on the local host latency. In netstacklat, we filter out any latency samples that may be inflated by TCP HoL blocking by default. In case it is relevant to include HoL delays, this filter can be disabled.
% Latency from HoL blocking can optionally be included, which may be useful in for example machines that only communicate within a data center network, where the issues causing HoL latency is still under the operator's control.

For the tcp-socket-enqueued probe, netstacklat itself does not need to perform an explicit exclusion, as the probed kernel function \texttt{tcp\_queue\_rcv()} only runs for in-order TCP segments. Any out-of-order TCP packets, i.e., those that may experience HoL blocking, take a separate path through the Linux network stack, where they are kept in an Out-Of-Order queue until they can be merged back in-order to the socket receive queue. However, for the tcp-socket-read probe, we need to take explicit action to avoid HoL blocked segments, as the read data may contain segments that originally arrived out-of-order and experienced HoL blocking.

To exclude measurements that may be inflated by HoL blocking from tcp-socket-read, we check if any new out-of-order packets have arrived at a socket since the last read from the socket. The kernel maintains a per-socket counter of out-of-order packets (\texttt{rcv\_ooopack}), so we can detect if new out-of-order packets have arrived by checking if the socket's counter has increased since the previous read. If new out-of-order packets have arrived, we compute the value $s_{max}$ (as detailed below) to be the maximum TCP sequence number that may arrive out of order. We then exclude reads that include sequence numbers below $s_{max}$ from the latency measurements. We store $s_{max}$ as part of the socket (\texttt{BPF\_MAP\_TY\-PE\_SK\_STO\-RAGE}) to persist for future reads, and invalidate it once the reads are ahead of $s_{max}$ to avoid inadvertently applying it after a wrap-around in the sequence range. 

As we only detect the presence of new out-of-order packets by the time the application reads from the socket, there are two possibilities that we have to consider when computing $s_{max}$. 1) Out-of-order packets are still in the out-of-order queue, in which case $s_{max}$ lies in the tail of this queue, or 2) The out-of-order packets have already been merged in-order to the socket's receive queue. For 1), Linux conveniently keeps an \texttt{ooo\_last\_skb} pointer in the socket data structure, which points to the tail of the out-of-order queue, so we simply retrieve $s_{max}$ as the sequence number of that last segment. For 2), the socket's next expected sequence number (\texttt{rcv\_nxt}) must have advanced past the out-of-order segments, and we therefore set $s_{max}$ to \texttt{rcv\_nxt}.

This procedure to exclude latency samples from potentially HoL blocked reads is deliberately conservative, and may also exclude some measurements taken after the HoL blocking has been resolved. However, because the HoL blocking is caused by network conditions outside the host itself, we consider this a reasonable tradeoff to avoid HoL blocking artificially inflating latency values, and thus obscuring the delays actually caused by the host behavior. Furthermore, we expect the fraction of packets that arrive out-of-order, either due to packet loss or packet reordering, to be relatively low for most systems. Therefore, the fraction of samples impacted by this mechanism should be minor.

\section{Evaluation in testbed}
\label{sec:workload-comparison}

\begin{figure}
    \centering
    \includegraphics[width=\columnwidth]{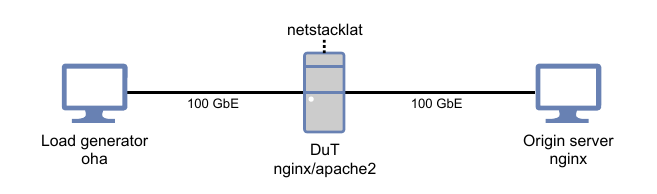}
    \Description{Schematic of the 3-node testbed setup, with a load generator on the left, the device under test in the middle, and an origin server on the right.}
    \caption{Testbed setup}
    \label{fig:testbed}
\end{figure}

To evaluate the monitoring capabilities and overhead of netstacklat, we assess its performance for several Nginx and Apache server workloads. 
% In Section~\ref{sec:testbed-setup} we describe the testbed setup and the workloads used, in Section~\ref{sec:workload-latency} we dissect the network stack latency reported by netstacklat, in Section~\ref{sec:workload-correlation} we compare the network stack latency to the end-to-end performance, and in Section~\ref{sec:workload-overhead} we summarize netstacklat's monitoring overhead. 
We dissect the network stack latency reported by netstacklat, compare the network stack latency to the end-to-end performance, and summarize netstacklat's monitoring overhead. Before diving into our experimental results, we introduce the testbed setup and workloads used.
We have made the measurement data, relevant configuration files, and scripts for running the experiments and plotting the results presented in this section and Section~\ref{sec:tool-comparison} publicly available\footnote{The dataset is available at \url{http://doi.org/10.5281/zenodo.19483855} and the scripts and configuration files at \url{https://github.com/simosund/netstacklat-imc26-scripts/}.}.

\subsection{Experimental setup}
\label{sec:testbed-setup}
We set up a testbed as illustrated in Figure~\ref{fig:testbed}, where netstacklat is used to monitor a server that acts as the Device under Test (DuT). We use a simple chain topology, where the load generator is directly connected to the DuT. The DuT is, in turn, also connected to an additional machine that acts as the origin server (running Nginx) for scenarios where the DuT is configured as a reverse proxy. All machines are equipped with Mellanox ConnectX-5 dual-port 100GbE NICs, providing enough network capacity to shift the bottleneck to the DuT host network. The DuT is equipped with a 6-core Intel Xeon E5-1650 CPU, while the load generator has a 10-core 20-thread Intel Xeon Silver 4114 CPU. Both machines run a Linux kernel 6.8 and have 32 GB of RAM.

To evaluate how netstacklat performs across several different workload patterns, we evaluate 4 different server configurations: Nginx as a server, Apache (apache2) as a server, Nginx as a reverse proxy, and Apache as a reverse proxy. We note that the objective of these experiments is to evaluate netstacklat across different workloads, not to make a performance comparison between Nginx and Apache. We therefore use the default Nginx and Apache configurations packaged for Ubuntu 24.04 instead of attempting to optimize and equalize their configurations. For the reverse proxy configurations, we disable caching, so each request fetches the resource from the origin server. While it would be more efficient for reverse proxies to cache static content, disabling caching creates a network load that is more distinct from that of the ordinary server configuration. 
For each of the 4 server configurations, we also test 4 different file sizes: 1kB, 10kB, 100kB, and 1MB.

We use the HTTP load generator oha~\cite{sugitaOha2025} to request the static files from the DuT. As the sustainable request rate varies greatly depending on server configuration and the size of the requested file, we instead vary the number of concurrent connections to create different levels of load, generating requests back-to-back on each connection. We vary the number of concurrent connections between 1, 5, 10, 50, 100, 500, 1000, 2000, and 4000 connections.

With 4 different server configurations, 4 different file sizes, and 9 different levels of concurrent connections, we get 144 different HTTP workload combinations. Each workload test runs for 60 seconds and is repeated 9 times, unless otherwise mentioned.

% Potential TODO: Should probably also mention how netstacklat was configured somewhere... E.g. running with ebpf-exporter (collecting stats every 5 seconds), using filtering and grouping on both interface and cgroup, essentially utilizing all of netstacklat's features and maximizing its overhead (as close to 100% of the traffic will be included by the fitlers, only excluding a minor amount of SSH traffic and Prometheus scrape resquests on the management interface)

\subsection{Latency distribution}
\label{sec:workload-latency}

\begin{figure}
    \centering
    \includegraphics[width=\columnwidth]{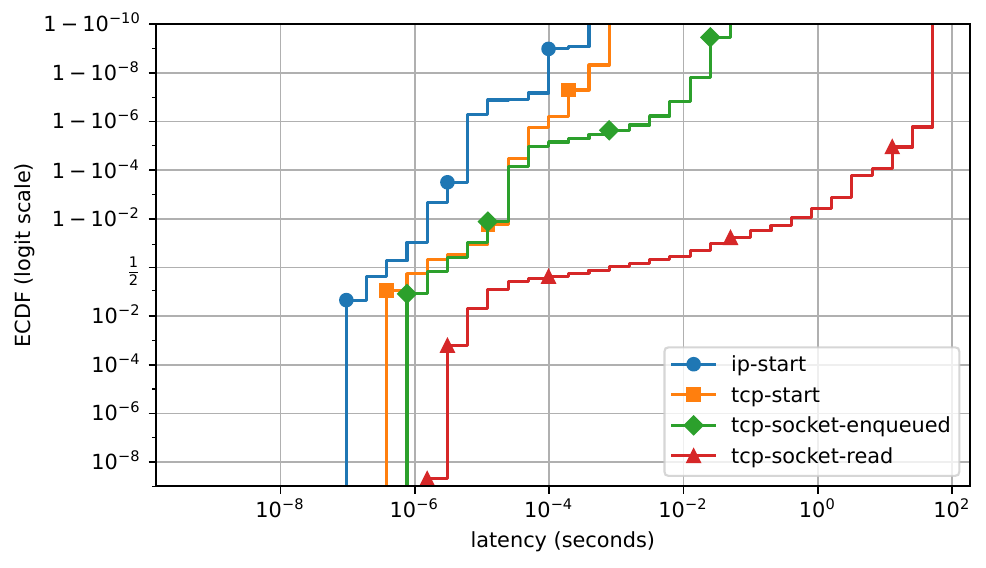}
    \Description{An emperical CDF showing the distribution of the latency for the 4 probe points. The scales highlight the very long latency tails, which are several orders of magnitude greater than the typical latency.}
    \caption{Joint latency distribution across all 144 tests.}
    \label{fig:alltests_ecdf}
\end{figure}

We first provide an overview of the latency measurements from netstacklat. Note that while we can calculate the exact mean latency from the sum and counts netstacklat keeps, for quantiles (including the median), we can only determine in which histogram bin they lie. In the figures, we use the middle point of the bin as the quantile estimate, which minimizes the maximum relative error
%, and for our power-of-two exponential histograms it translates to a possible relative error of 
to $\pm33$\%. E.g., if the quantile is in the $\left(2\text{us}, 4\text{us}\right]$ bin, we will show it as 3us. When discussing the results, we will refer to the bin edges rather than the displayed midpoint, and the conclusions we draw from the results would not change regardless of where within a bin the true quantile lies. 
% Although this may seem like a large error margin, the conclusions we draw from the results in this section would not change regardless of where within the bins the true quantile values lie.    

Figure~\ref{fig:alltests_ecdf} shows the joint latency distribution of all 144 workloads for each of the 4 netstacklat probe points in the TCP data path. Note that we use a logit scale on the y-axis to highlight the latency tail. Here we can see that all four probe points show a wide range of possible values. Indeed, the spread of values within each probe point is wider than the distance between them, i.e., the highest latencies for ip-start exceed the lowest latencies for tcp-socket-read.
% 1 - 0.99968563 of ip-start > 4us
% 6.06155887e-04 of tcp-socket-read <= 4us
% -> top 0.031\% of latencies for ip-start are higher than the bottom 0.061\% of latencies for tcp-socket-start. 
Even though the testbed has a relatively simple network setup with no traffic classifiers or VLAN tags that need to be processed before the IP-layer, we find that already for the early ip-start probe, the maximum latency reaches 262 - 524us, which is at least three orders of magnitude larger than the minimum observed value. 

For later probe points, the latency tail grows even longer and wider. For tcp-socket-enqueued, the observed latency spans more than four orders of magnitude, with roughly 1 in every 10 million samples exceeding 8ms. In extreme cases, a packet may even spend more than 33ms traversing the kernel network stack up to the socket, similar to the full end-to-end latency over a satellite network~\cite{mohanMultifacetedLookStarlink2024, garciaDetailedCharacterizationStarlink2025}. % e.g. "A Multifaceted Look at Starlink Performance" shows minimum Starlink latencies < 25ms
The total latency until the application actually reads the data, shown by tcp-socket-read, has the largest latency tail, with observed latency differences that span over seven orders of magnitude. Due to the tested workloads frequently overloading the DuT, roughly half of all reads have a host latency of 1ms or more, and around 0.37\% of the latencies exceed a whole second.

\begin{figure}
    \centering

    \subcaptionbox{Mean, median and 99th percentile latency for all 4 hooks\label{fig:alltests_latency_summary}}{
        \includegraphics[width=\columnwidth]{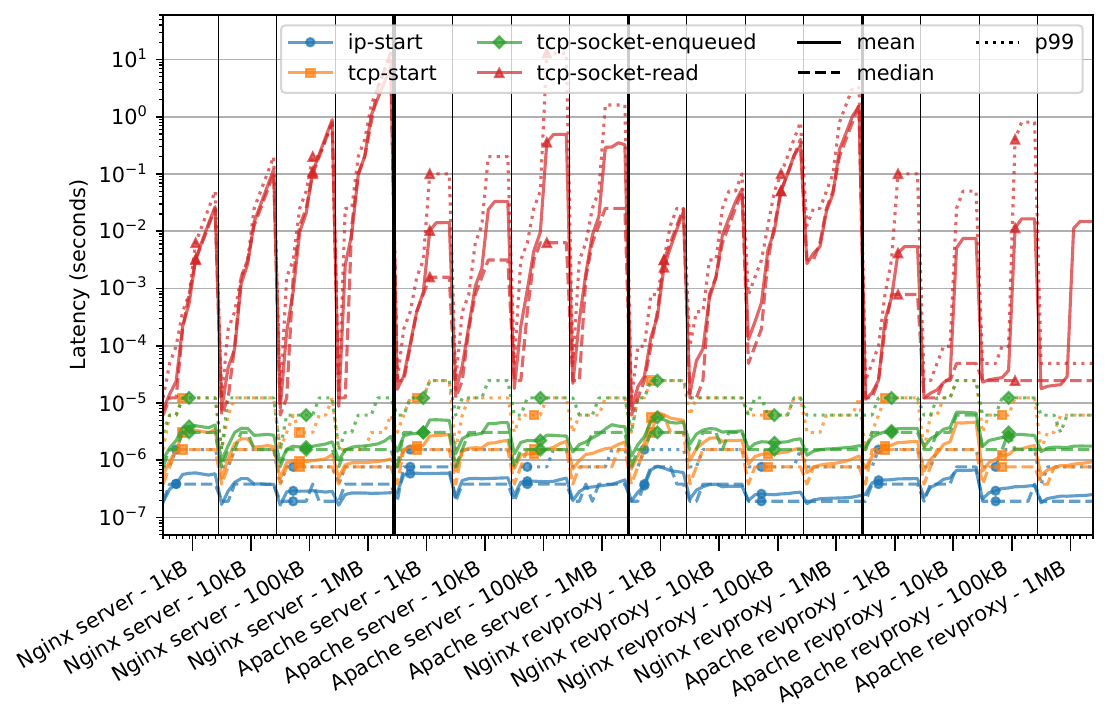}
        \Description{Line-graph showing how mean, median, and 99th percentile latencies the 4 probe points evolve across all 144 workloads. A clear saw-tooth pattern shows how the latency increases with an increasing number of parallel connections for all 16 server configuration and file-size combinations.}
        % Potential TODO: Add in the end-to-end latency as a 5th "probe point" in this figure, although that gets very messy
    }
    \subcaptionbox{Latency distribution for tcp-socket-read\label{fig:alltests_tcpread_latency}}{
        \includegraphics[width=\columnwidth]{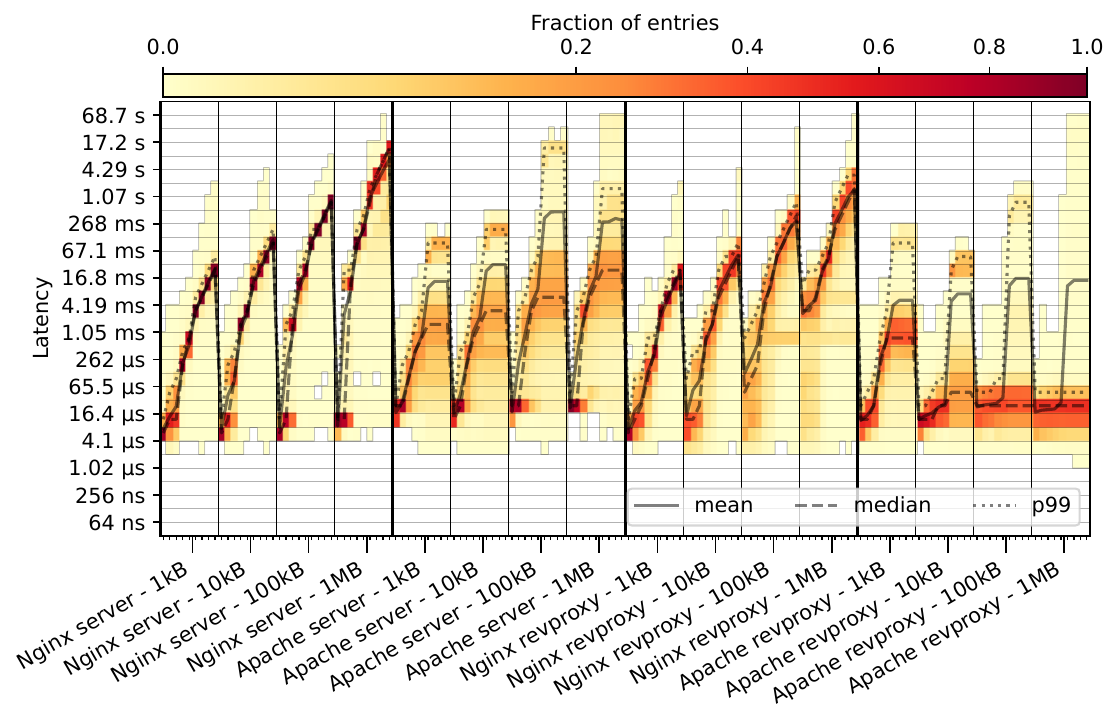}
        \Description{A heatmap showing the tcp-socket-read latency distribution for all 144 workloads. The distribution for Nginx shows a very narrow peak for each workload, whereas the latency for Apache has more variation.}
    }
    \caption{Latency (y-axis) for each of the 144 workload combinations (x-axis). The thick vertical lines separate the 4 server configuration groups. The thinner vertical lines and x-ticks group the results for each of the 4 file sizes within each server configuration. Within each of the 16 server-configuration and file-size combinations, the 9 levels of concurrent connections (minor ticks) are sorted in ascending order.}
    \label{fig:alltests_latency}
\end{figure}

To understand how latency varies between the different workloads, we break down the latency for each of the 144 workloads in Figure~\ref{fig:alltests_latency}. 
%We show the 16 server configuration and file size combinations on the x-axis, and within each of the 16 groups the 9 levels of concurrent connections are sorted in ascending order. 
Figure~\ref{fig:alltests_latency_summary} shows the mean, median, and 99th percentile of the latency distribution for each probe. Here we see that the latency typically grows at all 4 probe points as the load increases from a single connection up to 100 connections (the left-most part in each of the 16 x-axis groups). Several previous works~\cite{caiKernelVsUserlevel2023, awamotoOpeningKernelBypassTCP2025} have noted that many concurrent connections increase the latency of the Linux network stack, partly due to reduced cache hit rates. Here, it is apparent that even the early parts of the network stack running before ip-start take several times longer to complete when there are many concurrent connections. Past 100 connections, however, the latency increases for ip-start, tcp-start, and tcp-socket-enqueued mostly flatten out, while the latency until the application actually reads the data at tcp-socket-read typically continues to grow as the request backlog piles up. The latency inflation as the number of connections increases therefore ends up being much more modest for ip-start, tcp-start, and tcp-socket-enqueued, which stay within one order of magnitude of their single-flow latency. In contrast, the application latency with tcp-socket-read is typically inflated by three orders of magnitude or more.

% Potential TODO: Also comment that the 99th percentile for ip-start, tcp-start and tcp-socket-enqueued even at a single flow typically exceeds the median and mean latency at 1000+ flows, i.e. the tail in unloaded conditions is longer than the increase due to load.

We further break down the latency for tcp-socket-read by showing the full histogram distribution in Figure~\ref{fig:alltests_tcpread_latency}. Here we can observe that while all four server configurations show a large increase in latency with more concurrent connections, the latency distributions differ considerably between them. Both as a server and as a reverse proxy, Nginx shows a consistent increase in latency as the number of connections increases to 4000. For Apache on the other hand, the latency flattens out past 1000 concurrent connections. This is due to the default settings supporting fewer concurrent connections for Apache, so it starts to drop requests when there are many concurrent connections, instead of building up a long request backlog like Nginx. %(Nginx can be considered to suffer from more "request bufferbloat").
Furthermore, Nginx shows fairly consistent latency for each read, where the majority of all reads within a workload combination experience similarly high latency. In contrast, as Apache drops some requests and uses a separate thread for each request, it sees a much more spread-out latency distribution. Apache always has a fraction of reads that complete relatively fast, and likewise, only a fraction of the reads experience the highest multi-second latencies.

Apache as a reverse proxy serving files that are 10kB or larger, also stands out from the rest, as the vast majority of the reads experience a host latency below 65us here. This is due to the bottleneck shifting from the DuT to the origin server in these scenarios, 
% as the Apache reverse proxy configuration causes the origin server to compress the result (HTTP 1.1), while for the Nginx reverse proxy the origin server sends an uncompressed response (HTTP 1.0) that the DuT then compresses
and therefore CPU contention no longer causes high latencies. However, there is still a minor fraction of reads that experience high latencies; in the case with 1MB files, some even exceed 8 seconds. While not directly visible in Figure~\ref{fig:alltests_tcpread_latency}, latencies above 1ms occur only on the interface facing the load generator, not on the interface towards the origin server (see Figure~\ref{fig:apdix_workloads_tcpread_dist_pernic} in the appendix). The high latency values are in this case caused by some requests still needing to wait a long time before an Apache worker can start processing them, as Apache still needs to wait on the origin server before it can complete ongoing requests.

\subsection{Correlation to end-to-end metrics}
\label{sec:workload-correlation}

% Potential TODO: Discussion around how much of the end-to-end latency that the tcp-socket-read latency reflects. Could use e.g. Figure~\ref{fig:e2e_fraction} or~\ref{fig:alltests_e2e_vs_tcpread_latency} (see appendix), although they get a bit strange due to the quantile estimation errors of netstacklat's power-of-two histograms causing the network stack latency to at times "exceed" the end-to-end-latency.

\begin{figure}
    \centering
    \includegraphics[width=\columnwidth]{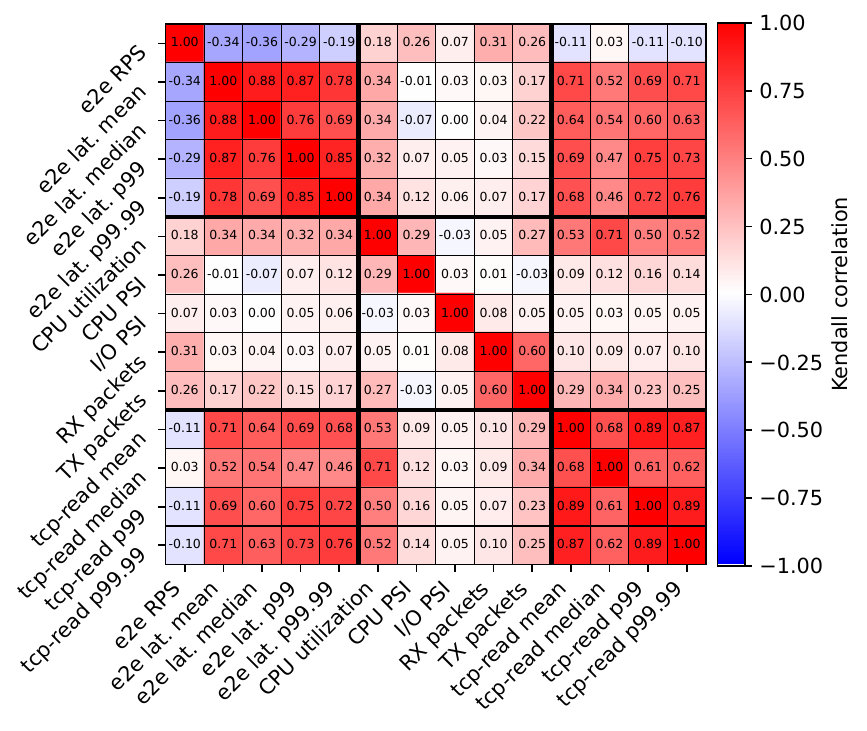}
    \Description{A correlation matrix showing the Kendall correlation between the end-to-end delay, DuT system load, and DuT tcp-socket-read latency. Tcp-socket-read latency shows a strong correlation (0.5-0.7) with the end-to-end delay, while CPU load only has a moderate correlation (0.3).}
    \caption{Kendall correlation between end-to-end (e2e) latency, system load, and tcp-socket-read latency for reading the requests.}
    \label{fig:correlation_matrix}
\end{figure}

We further examine how well the tcp-socket-read latency for reading the requests on the DuT correlates with the end-to-end request latency reported by the load generator. Figure~\ref{fig:correlation_matrix} shows the Kendall correlation between three groups of metrics: end-to-end RPS and request latency reported by the load generator; system load metrics for the DuT, such as CPU utilization, CPU PSI~\cite{weinerPSIPressureStall2018}, and packet rates; and tcp-socket-read latency for the DuT interface where requests arrive. Here, we find that tcp-socket-read latency strongly correlates with end-to-end request latency. The tcp-socket-read median latency appears to have a weaker correlation to the end-to-end delay than the tail latency, indicated by the 99th percentile, and the mean, which is also impacted by the tail latency. Of the system load metrics, CPU utilization has the strongest correlation to the end-to-end latency, although it is much weaker than the correlation between tcp-socket-read and end-to-end latency.

We note that our testbed experiments are set up in a manner that strongly benefits the correlation for tcp-socket-read and limits the correlation for the system metrics. Our load generator is directly connected to the DuT on a link with ample bandwidth and minimal latency, so that the host latency becomes the primary source of latency for the end-to-end network connection. For servers connected to more complex networks where network latency (outside the host) plays a significant role, the correlation between end-to-end latency and host latency will likely be significantly lower. Furthermore, many of our workload combinations intentionally overload the DuT to cause host congestion. This limits the potential correlation for CPU utilization, as it will frequently be close to 100\% and unable to distinguish the degree to which the system is overloaded. In real server deployments where load balancing and resource scaling techniques are employed, the CPU utilization will typically not reach 100\%, and we expect that it will then have a stronger correlation to both host latency and end-to-end performance. Nonetheless, host network latency, and especially tail latency, remains a promising signal to detect performance issues. Additionally, it may also be useful as a load indicator to consider when making load balancing and resource scaling decisions.

% \subsection{Kernel jitter} % Unlikely to have much time for this
% Just an idea, but could be interesting to also use netstacklat to monitor the latency during a very low load scenario and then e.g. tweak some power settings or play around with CPU affinity (having some different process running on the same vs pinned to different CPUs) to see how the tail latency is affected. This might actually be more a more interesting showcase for latency monitoring than the massively overloaded server scenario.

\subsection{Overhead}
\label{sec:workload-overhead}

\begin{figure}
    \centering
    % \includegraphics[width=\columnwidth]{figures/netstacklat_alltests_cpu_relsys_allprogs_stacked.pdf}
    % \caption{CPU utilization for netstacklat relative to system load.}
    \includegraphics[width=\columnwidth]{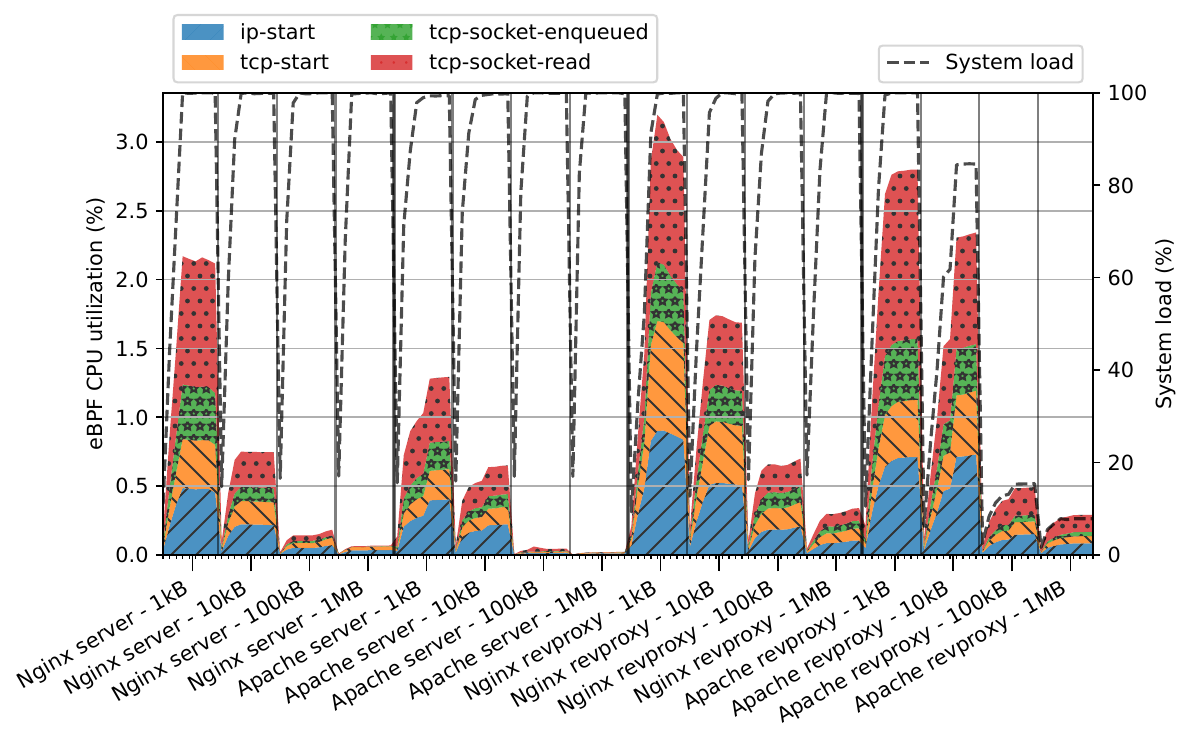}
    \Description{A stacked area-graph showing the overhead for each of the 4 netstacklat probe points across all 144 workload combinations, with the combined overhead varying between less than 0.1 percent to just over 3 percent CPU utilization. The overhead is highest for the reverse proxy configurations with smaller file sizes, with a decreasing trend for larger file sizes.}
    \caption{Average CPU utilization for netstacklat eBPF programs for each of the 144 workloads. Same x-axis layout as Figure~\ref{fig:alltests_latency}.}
    \label{fig:alltests_cpu_perhook}
\end{figure}

To measure the overhead imposed by netstacklat, we use the cumulative eBPF runtime and invocation counts reported by the Linux kernel via the \texttt{bpf\_stats\_en\-abled} setting. From these, we can infer the average overhead of netstacklat's programs, but not the per-invocation variability.
Figure~\ref{fig:alltests_cpu_perhook} shows the average CPU utilization for all netstacklat's eBPF programs for each of the 144 workloads, broken down by the 4 netstacklat probe points in the TCP path. The dashed line shows the total system CPU utilization during each load test on the right y-axis, which typically reaches 100\% when there are 50 or more concurrent connections. As can be seen, the overhead varies greatly between workloads. For the Nginx reverse proxy setup serving 1kB files, netstacklat reaches upward of 3.2\% CPU utilization, while less than 0.1\% CPU utilization is needed when monitoring Apache serving 1MB files.

% Potential TODO: Comment on the breakdown of overhead for the different programs/probe-points. Overhead is highest for tcp-socket-read as it does cgroup filtering (which includes an additional hash map lookup) and has additional logic to exclude HoL-blocked reads.

In general, requests for smaller file sizes result in higher monitoring overhead, as they lead to higher request rates and thus more packets to monitor. Similarly, the reverse proxy configurations also result in higher monitoring overhead due to additional ingress traffic from the origin server. We note, however, that for the Apache reverse proxy workloads, the overall system load decreases at large file sizes as it shifts the bottleneck over to the origin server. While netstacklat only reaches about 0.3\% CPU utilization for the Apache reverse proxy setup with 1MB files, that makes up about 3.7\% of the system's total 7.8\% CPU utilization for that workload. On average, across the 144 workload combinations, netstacklat had an absolute overhead of 0.81\% CPU utilization, or a relative overhead of 1.37\% of the system's total CPU utilization. 

Although this overhead is not insignificant, we argue that it is low enough to make continuous monitoring feasible in many scenarios. For web servers in the wild, there is often significantly more work involved with handling requests than the minimal setup used to serve static files in our testbed. The relative overhead from netstacklat can therefore be expected to be even lower than in our testbed, as illustrated through the Cloudflare deployment in Section~\ref{sec:cdn-exp-loadincrease}. However, we also acknowledge that as the overhead scales with the packet rate, the overhead might still pose a challenge for systems where the amount of ingress traffic is very high relative to the system's compute workload. For such high-traffic scenarios, sampling the measurements could help reduce the overhead, as further discussed in Section~\ref{sec:limitations}.
% Thanks to the independent nature of netstacklat's probe points and latency calculation, overhead can be further reduced by disabeling probe-points that are not of interest, or employ sampling

% Potential TODO: Should maybe have some plot with the request rate or ingress packet-rate to put the overhead into perspective of something else than just CPU utilization.
% - This is in the appendix for now

\section{Comparing overhead of monitoring tools}
\label{sec:tool-comparison}
We now look at how the overhead from netstacklat compares to several of the state-of-the-art tools presented in Section~\ref{sec:related-work}. 
For brevity, these comparisons will focus on a single workload scenario from Section~\ref{sec:workload-comparison}: an Nginx server serving 10kB files. This is roughly the median option in terms of overhead from Section~\ref{sec:workload-overhead}.
Although other workload scenarios will result in different levels of absolute overhead, less extensive testing with other workloads from Section~\ref{sec:workload-comparison} has been performed to verify that the relative performance difference between the tools is similar. % TODO - add reference to appendix here (and add the "multitest" plots there)
% we expect that the relative performance difference between the monitoring tools will be similar, as the overhead scales proportionally to the network load.
We will first describe our tool selection and configuration, then look at the probing overhead from the tools' eBPF programs, and finally show how the overhead impacts the end-to-end performance.
% In Section~\ref{sec:tool-config} we cover how we configure the different tools, while Section~\ref{sec:tool-cpu-util} shows the overall CPU overhead from the tools and Section~\ref{sec:tool-e2e-impact} looks at how the tools impact the end-to-end performance.

\subsection{Tool configuration}
\label{sec:tool-config}

We consider all publicly available tools from Section~\ref{sec:related-work} that can be used for our HTTP workloads from Section~\ref{sec:workload-comparison}. 
The set of tools we compare netstacklat against thereby includes lattrace~\cite{maurerInvestigatingCausesJitter2021}, kyanos~\cite{kyanos-webpage}, retis~\cite{tenartChallengesLimitationsDebugging2025}, and pwru (packet where are you)~\cite{liangPwruLinuxKernel2024}. In addition, we also perform the tests without any monitoring tool to establish a baseline. 

For lattrace, there are no relevant options to modify. Kyanos, on the other hand, has several different monitoring modes. We use kyano's stat command to allow it to aggregate the results instead of outputting individual traces, as netstacklat only provides aggregated data. Furthermore, we specified \enquote{http} as the L7 protocol for kyanos to monitor. 
% While lattrace and kyanos, similarly to netstacklat, use their own pre-defined set of kernel functions to probe, both retis and pwru can be configured to probe any number of kernel functions that receive SKBs as arguments.
Unlike netstacklat, lattrace, and kyanos, which all use their own pre-defined set of kernel functions to probe, retis and pwru can be configured to probe any number of kernel functions receiving SKBs as arguments. For retis we use their \enquote{generic} profile, which probes 25 kernel functions, while pwru, by default, probes all kernel functions that process SKBs. As the overhead from all probes adds up, retis and pwru can be expected to have higher overhead than netstacklat simply due to their greater number of probes rather than architectural differences. We therefore also test retis and pwru configured to only monitor the same kernel functions that netstacklat does, which we refer to as retis-netstacklat and pwru-netstacklat, respectively. However, note that netstacklat also probes \texttt{tcp\_recv\_timestamp()}, which neither retis nor pwru can monitor. 
% as that function does not receive an SKB as an argument
% (at least without adding a bunch of extra functionality in retis/pwru that will add overhead and thereby counteract the whole point of reducing their probe count to lower their overhead)

% For all tools except netstacklat, we direct all output to \texttt{/dev/null} to prevent their comparatively large amount of output from creating I/O spikes.
% We also restart the monitoring applications between each test, as lattrace and retis risk running out of memory otherwise

\subsection{CPU overhead}
\label{sec:tool-cpu-util}

\begin{figure}
    \centering
    \subcaptionbox{CPU utilization\label{fig:tools_cpu_util}}{
        \includegraphics[width=0.95\columnwidth]{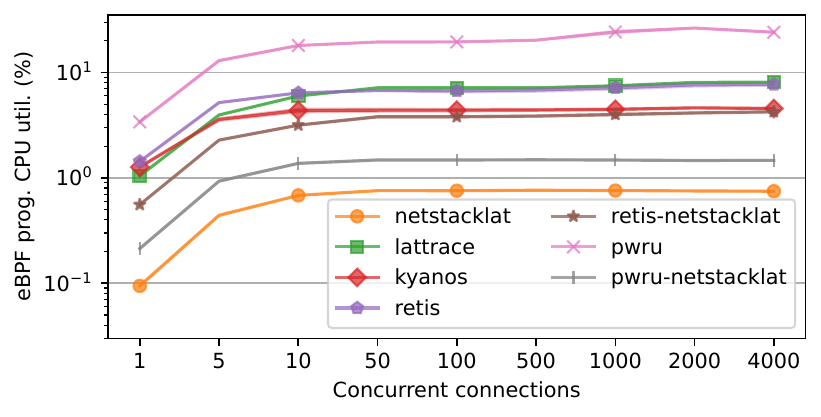}
        \Description{Line-graph showing the CPU utilization for the eBPF programs of the different tools. The CPU utilization increases with the number of concurrent connections up to 50 connections, and then flatten out. Netstacklat tops out around 0.75 percent CPU utilization, while pwru-netstacklat being second best with 1.5 percent CPU utilization, and other tools being drastically higher.}
    }
    \subcaptionbox{Average runtime\label{fig:tools_runtime}}{
        \includegraphics[width=0.95\columnwidth]{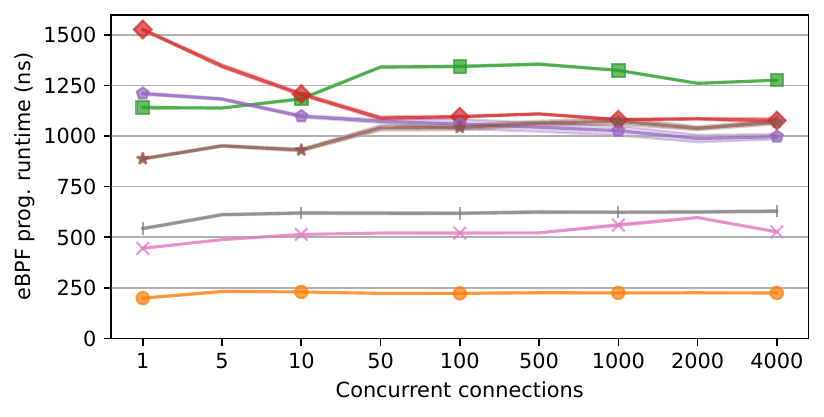}
        \Description{Line-graph showing CPU time spent per invocation of each of the tool's eBPF programs. For most tools, the runtime remains similar regardless the number of concurrent connections. Netstacklat has the lowest runtime at 225ns, while the second-best is pwru with just over 500ns.}
    }
    \caption{Overhead from eBPF programs of monitoring tools. The line shows the mean across all repetitions and the (narrow) shaded area shows the 95\% confidence interval.
    %A shaded area around the lines shows the 95\% confidence interval, although it is very narrow relative to the scale of the plots.
    }
    \label{fig:tools_overhead}
\end{figure}

Figure~\ref{fig:tools_overhead} shows the CPU overhead from the eBPF programs based on the kernel's eBPF runtime performance counters. As the baseline has no overhead, it is not included here. In Figure~\ref{fig:tools_cpu_util}, we can see that once the DuT is fully loaded from handling 50 or more concurrent connections, netstacklat stabilizes around 0.75\% CPU utilization. This is roughly half of the 1.5\% for pwru-netstacklat, and only about 15\% of the overhead compared to kyanos. While pwru has relatively low overhead when probing the same functions as netstacklat, we can also note that, in its default mode of probing every function processing SKBs, the overhead from pwru's eBPF programs reaches upwards of 25\%. It is thereby clear that probing every kernel function in the network stack comes at a high overhead cost, so while highly useful for debugging and understanding the network stack, it is not a feasible solution for continuous monitoring.

\begin{figure*}[ht!]    
    \centering
    \subcaptionbox{RPS\label{fig:tools_e2e_rps}}{
        \includegraphics[width=0.32\textwidth]{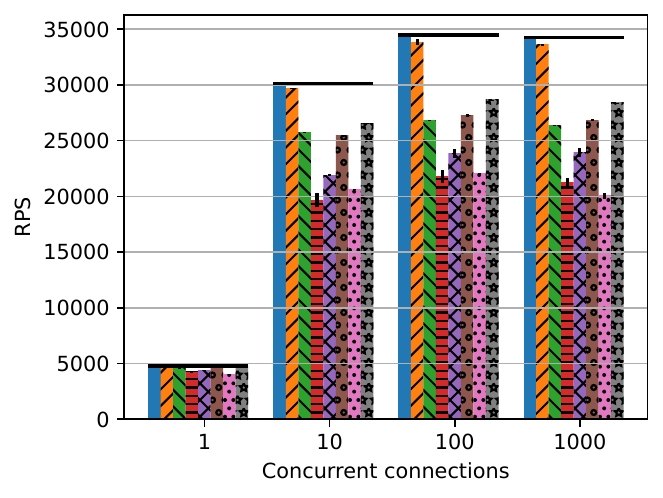}
        \Description{Bar plot showing achieved RPS without any monitoring (baseline) and with netstacklat and the other monitoring tools. While netstacklat only shows a marginal decrease (2 percent), all other tools show a substantial reduction in RPS past 10 concurrent connections (at least 17 percent).}
    }
    \subcaptionbox{Mean latency\label{fig:tools_e2e_latency_mean}}{
        \includegraphics[width=0.32\textwidth]{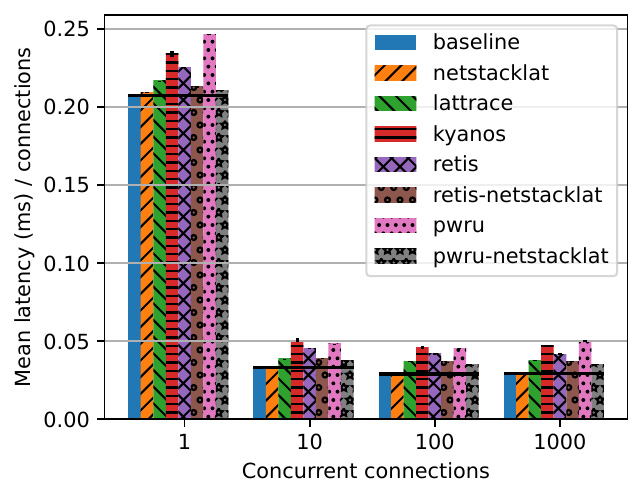}
        \Description{Bar plot showing the mean response time without any monitoring (baseline) and with netstacklat and the other monitoring tools. The tools show similar relative performance regardless of number of concurrent connections, with netstacklat being the only tool with no clear visual increase over the baseline (less than 2 percent).}
    }
    \subcaptionbox{99th percentile latency\label{fig:tools_e2e_latency_p99}}{
        \includegraphics[width=0.32\textwidth]{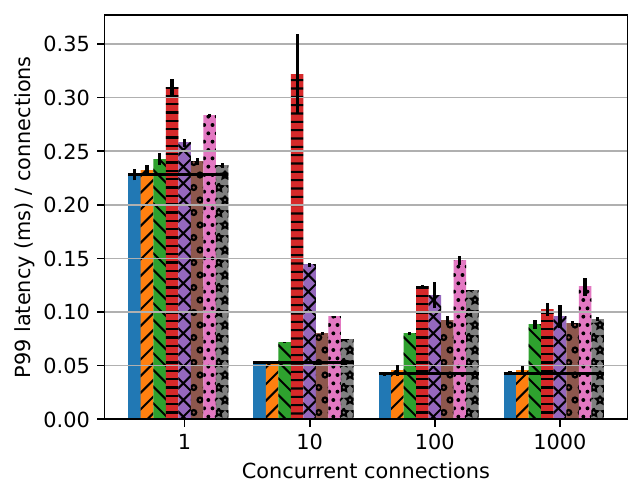}
        \Description{Bar plot showing the 99th percentile response time without any monitoring (baseline) and with netstacklat and the other monitoring tools. Most tools show a larger inflation (at least a 100 percent increase) compared to the baseline at 100 and 1000 connections, while netstacklat only has a slight increase over the baseline (6 percent).}
    }
    \caption{End-to-end impact of monitoring tools. The bar shows the mean across all repetitions, with the error bars indicating the 95\% confidence interval. The latency values in Figures~\ref{fig:tools_e2e_latency_mean} and~\ref{fig:tools_e2e_latency_p99} have been divided by the number of concurrent connections to provide a more convenient scale.}
    \label{fig:tools_e2e}
\end{figure*}

The comparatively low CPU overhead of netstacklat is largely explained by the much shorter runtime of its eBPF programs, shown by Figure~\ref{fig:tools_runtime}. Calculating the latency from the SKB timestamp and aggregating the value in a histogram only takes netstacklat between 220-230ns. On the other hand, the other tools have to push a timestamp with associated metadata to user space, which typically takes them just over 1us. Once again, pwru stands out as the best alternative to netstacklat, where its eBPF programs take around 520ns when probing every SKB function, about twice as long as netstacklat's. 
When pwru only probes the same functions as netstacklat, thus running less frequently and likely not utilizing the cache as efficiently, the average runtime increases to between 610-630ns -- almost three times as high as netstacklat.
%When pwru only probes the same functions as netstacklat and, thus, runs less often, the average runtime increases to between 610-630ns -- almost three times as high as netstacklat.

% Observation not relevant to the paper: Kinda interesting that pwru has significantly lower overhead per invocation than the other tools, despite using a similar model of pushing per-event messages. One noteable difference is that pwru uses a queue (BPF_MAP_TYPE_QUEUE) rather than a ring/perf buffer to send messages to user space. If I ever develop another eBPF programs that pushes messages to user space, I might have to try a queue instead of perf/ring buffer to see how that performs.

We note that the results in Figure~\ref{fig:tools_overhead} only consider the overhead of the monitoring tools' eBPF programs. For netstacklat, this is the primary concern when it comes to its scalability, as the eBPF programs need to run for every SKB. In contrast, netstacklat's user space agent only needs to periodically fetch the aggregated histogram data, and therefore, the overhead of the user space agent does not increase with the network traffic load. However, this is not the case for any of the other tools we compare against, which all process individual events on the user space side as well.

% Potential TODO (for revised version of paper): Measure the CPU overhead from the user space agent as well to strengthen these results.

\subsection{Impact on end-to-end performance}
\label{sec:tool-e2e-impact}
To provide a clearer picture of how the total overhead from the monitoring tools, including their user space components, impacts the
% overall performance of the 
DuT, we next look at the tools' impact on the end-to-end performance in Figure~\ref{fig:tools_e2e}. Figure~\ref{fig:tools_e2e_rps} shows the mean requests per second (RPS), while Figures~\ref{fig:tools_e2e_latency_mean} and~\ref{fig:tools_e2e_latency_p99} show the mean and 99th percentile request response latency reported by the HTTP load generator. 
% Note that the latency values in Figures~\ref{fig:tools_e2e_latency_mean} and~\ref{fig:tools_e2e_latency_p99} have been divided by the number of concurrent connections to provide a more convenient scale.
% For instance, the latency values at 1000 concurrent connections are 1000 times higher than shown. 
Under the lightly loaded scenario of a single concurrent flow, when there is plenty of CPU headroom for the monitoring tools to use, most of the tools perform reasonably well. Here, netstacklat only reduces RPS with 0.8\% and inflates P99 latency by less than 2\%. In comparison, the tools with the largest impact here, kyanos and pwru, reduce RPS by 11\% and 16\%, respectively, and simultaneously increase P99 latency by 36\% and 24\%, respectively.

However, as the load increases to 100 or 1000 flows, there is no longer any spare CPU headroom for the monitoring tools to use. Consequently, the impact on the end-to-end performance grows larger as the monitoring tools start to contended for the system resources used to handle the requests. In these heavily loaded scenarios, netstacklat reduces RPS with less than 2\%, while the other tools reduce RPS between 17 - 42\%. Similarly, netstacklat does not inflate the mean latency by more than 2\%, and P99 latency by no more than 6\%. In contrast, the other tools inflate the mean latency by between 20 - 71\%, and more than double the P99 latency. We conclude that, of the evaluated tools, netstacklat is the only tool suitable for continuous monitoring as it maintains a small footprint both during lightly and heavily loaded scenarios. The other tools are better suited for running in shorter periods of time when the system is not expected to be under heavy load or when considerable performance impacts are acceptable.

\section{Deployment at Cloudflare}
\label{sec:cdn-deployment}
In addition to the experiments on the testbed, we have also deployed netstacklat across all servers in Cloudflare's global CDN network. As Cloudflare already uses ebpf-exporter to collect other custom metrics, the deployment only required adding netstacklat's eBPF programs to the ebpf-exporter configuration. Some minor tweaks were made to adapt netstacklat to Cloudflare's environment, primarily adding compile-time flags to disable unneeded features. For this initial deployment, the primary metric of interest was the total network stack to application latency for TCP traffic, i.e., netstacklat's tcp-socket-read probe. To minimize overhead, all other probe points were disabled.
% The tcp-socket-read (and udp-socket-read) probe point is also the only probe point that supports filtering and grouping the data by cgroup, which was a desired feature to limit the monitoring to the applications of interest.
Furthermore, monitoring was limited to two services of interest (nginx-ssl and nginx-cache) by configuring netstacklat to filter and group the data by their cgroups. 

While the rollout at Cloudflare is recent (began 2025-11-05), we 
%in Section~\ref{sec:cdn-exp-loadincrease} and Seciton~\ref{sec:cdn-anomaly} 
below share initial insights from two scenarios. 
As the measurement data from this live deployment may contain operationally sensitive details, we unfortunately cannot share the raw data. 

% Updated: The percentiles are no longer based directly on Grafana's histogram_quantile, but use the same interpolation logic to produce nice looking results (rather than digital steps between each bin)
Like previously, the latency percentiles presented in this section are based on netstacklat's histogram data. However, rather than displaying the midpoint, we here linearly interpolate the percentile between the bin limits as done by Prometheus' \texttt{histogram\_quantile()} function. % https://github.com/prometheus/prometheus/blob/9a3ac8910b0476d0d73a5c36a54c55baec5829b6/promql/quantile.go#L105
This interpolation creates clearer timeline graphs, although the interpolated value may not accurately reflect where within the bin the true percentile lies. We will clarify the uncertainty of the percentiles when discussing the results.

% Potential TODO - Would maybe be good to have a brief section on overhead across the fleet (or at least one colo and/or generation of machines)
% - Partly done by adding a sentence about load the the busiest colo

\subsection{Latency under increasing server load}
\label{sec:cdn-exp-loadincrease}

\begin{figure}
    \centering
    \subcaptionbox{RPS. The dashed horizontal line shows the estimated capacity of the server (5000 RPS).\label{fig:cdn_expload_rps}}{
        \includegraphics[width=\columnwidth]{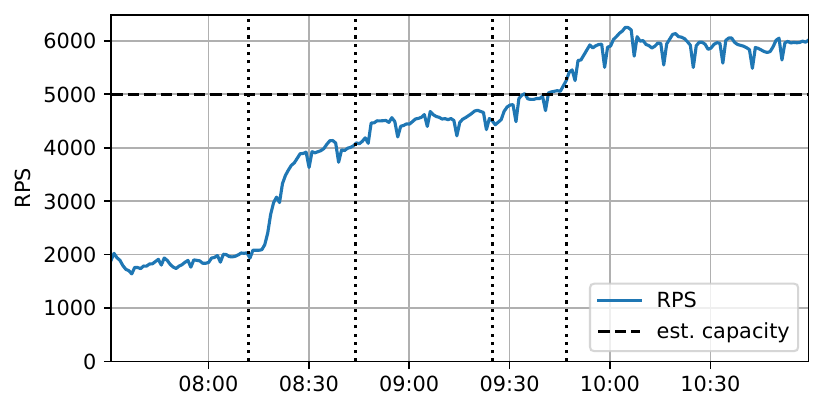}
        \Description{Timeseries-plot showing how the RPS handled by the server progressively increases as more traffic is directed towards it, starting around 2000 RPS and reaching 6000 RPS during the overloaded phase at the end of the experiment.}
    }
    \subcaptionbox{CPU utilization. The line shows the average across all cores, while the shaded area shows the span between the cores with the lowest and highest utilization.\label{fig:cdn_expload_cpu}}{
        \includegraphics[width=\columnwidth]{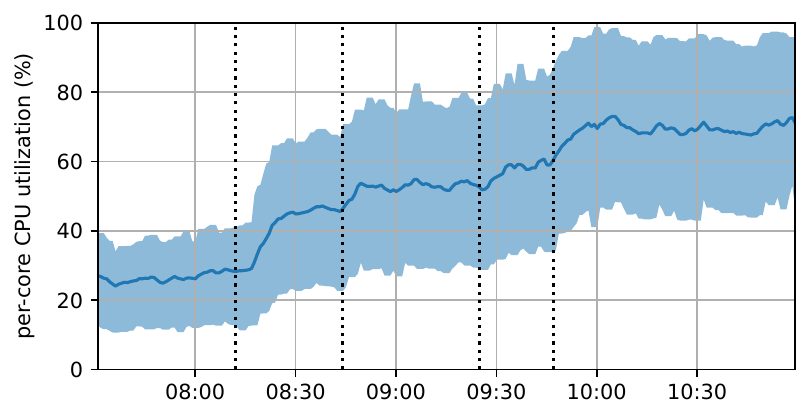}
        \Description{Timeseries-plot showing how the server's average CPU utilization increases proportionally to the RPS over the course of the experiment. A shaded area shows how the CPU utilization varies across the 192 cores, with some cores being close to 100 percent utilization during the overloaded phase of the experiment.}
    }
    \subcaptionbox{tcp-socket-read latency.\label{fig:cdn_expload_netstacklat}}{
        % There's also versions in log scales and with RPS on the secondary axis
        \includegraphics[width=\columnwidth]{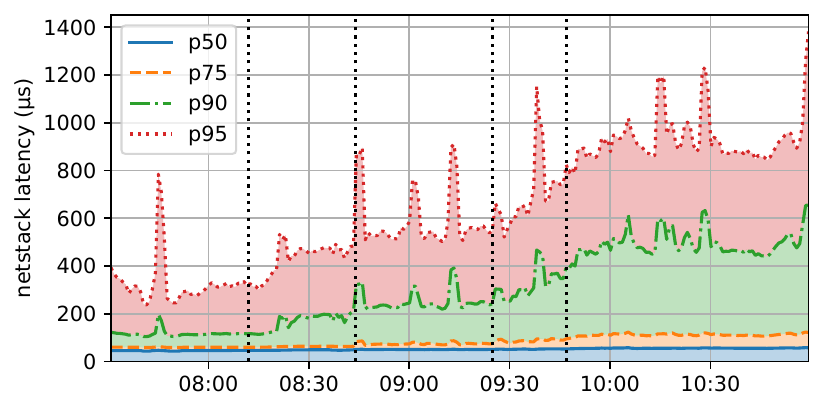}
        \Description{Timeseries-plot showing how the median, 75th, 90th, and 95th percentile tcp-socket-read latency increases over the course of the experiment. The median and 75th percentile only shows a slight increase, but the 90th and 95th percentile increases roughly proportionally with the RPS.}
    }
    \caption{Evolution of request rate, CPU load, and tcp-socket-read latency as the traffic load to a server is gradually increased. The four vertical dotted lines show when the load balancer was configured to increase the amount of traffic to the server to 4000, 4500, 5000, and 6000 RPS, respectively.}
    \label{fig:cdn_expload}
\end{figure}

One potential use case we foresee for netstacklat is to use it as a load signal, indicating when a server is becoming  overloaded. While utilization metrics can show a lack of spare resources, the latency before packets are processed by the application directly measures the consequence of the overload. We therefore perform a semi-controlled experiment to assess if netstacklat detects any significant latency increase for nginx-ssl as a server becomes overloaded, where Figure~\ref{fig:cdn_expload} summarizes the results. In this experiment, we configure the load balancer in front of a selected server to steer a gradually increasing amount of traffic towards the server. Figure~\ref{fig:cdn_expload_rps} shows how the request rate handled by the server increases during this process. During the last phase of the experiment, the load balancer is configured to serve 6000 RPS, intentionally creating a traffic load that exceeds the server's estimated 5000 RPS capacity. 

As the traffic load in Figure~\ref{fig:cdn_expload_rps} increases, the server's CPU load in Figure~\ref{fig:cdn_expload_cpu} shows a proportional increase. Once the traffic load reaches 6000 RPS, the average CPU utilization reaches around 70\%, although some cores are close to 100\% utilization, suggesting that the traffic load is pushing the server to its limits. The tcp-socket-read latency recorded by netstacklat in Figure~\ref{fig:cdn_expload_netstacklat} shows the direct impact the increased load has on the tcp-socket-read latency for nginx-ssl. The interpolated median latency shows a small increase from roughly 45us to 55us, but this still falls within the same histogram bucket, so we cannot be certain that the true median has increased. However, the tail latency in the form of the 90th percentile shows a much more significant inflation, increasing from around 100us (in the 64-128us bucket) before the start of the experiment to around 500us (shifting between the 256-512us and 512-1024us buckets) at 6000 RPS. We can thereby conclude that the increased traffic load has a clear impact on the tail of the host network latency.

While the tcp-socket-read latency did substantially increase during the overloaded phase of the experiment, 
%the increase was not as drastic as we had initially expected. Tail latencies
tail latency in the sub-millisecond range is still unlikely to have a major effect on the total end-to-end performance. We found that the nginx-ssl service initially receiving the requests was not the bottleneck in this experiment. Instead, Cloudflare's FL system~\cite{cloudflare-FL2-blog}, which handles the request after it has been received by nginx-ssl, was the primary bottleneck. Internal metrics showed that during the overloaded phase, the time FL spent handling the requests increased considerably and was many times greater than the tcp-socket-read latency for nginx-ssl. 
% Further increasing the traffic load would therefore not necessarily be problematic for the nginx-ssl service. However, if nginx-ssl has to stall the reception of new requests to wait on FL to complete prior requests, this could still results in high tcp-socket-read latency, similarly to Apache reverse proxy scenario in Section~\ref{sec:workload-latency}.
However, as Cloudflare transitions to the new FL2 system~\cite{cloudflare-FL2-blog} and test different CPU allocation strategies, the bottleneck could shift away from FL. By comparing the tcp-socket-read latency and FL's request service time, Cloudflare can now better identify where requests spend the majority of their time.

\begin{figure}
    \centering
    \includegraphics[width=\columnwidth]{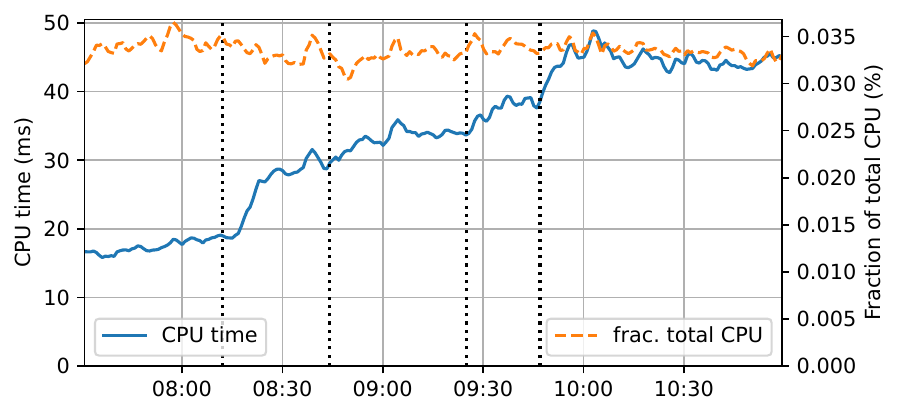}
    \Description{Timeseries-plot showing how the CPU time spent by netstacklat increases roughly proportionally with the RPS (from 20ms to 45ms per second), but relative to the overall CPU utilization it stays constant (around 0.035 percent).}
    \caption{CPU time spent by netstacklat's eBPF program (left axis) and the corresponding fraction of the server's total CPU utilization (right axis).}
    \label{fig:cdn_expload_overhead}
\end{figure}

In terms of overhead, Figure~\ref{fig:cdn_expload_overhead} shows the average CPU time per second consumed by netstacklat throughout the experiment. The overhead increases proportionally to the traffic load, as it scales with how frequently the probed TCP sockets are read, on average adding 225ns per read. Even during the overloaded phase with 6000 RPS, netstacklat only consumes around 45ms of CPU time per second, corresponding to 4.5\% utilization of a single CPU core (on a 192-core server). The dashed line compares netstacklat's CPU overhead with the server's total CPU utilization. As the server's overall CPU load also increases with the traffic load, the fraction of the total CPU load caused by netstacklat remains around 0.035\% throughout the experiment. 
% Potential TODO - Make script for Jesper to gather this type of data for a colo during e.g. a week. Could have CDF of CPU-time, or CPU-time normalized with total system load
At the busiest data center location, the daily peak of the average CPU time across all servers was around 85ms, or correspondingly 0.04\% CPU utilization. The overhead added by netstacklat is hence very small compared to the overall system load.

\subsection{Temporary latency anomaly}
\label{sec:cdn-anomaly}

\begin{figure}
    \centering
    \subcaptionbox{tcp-socket-read latency\label{fig:cdn_anomaly_latency}}{
        \includegraphics[width=\columnwidth]{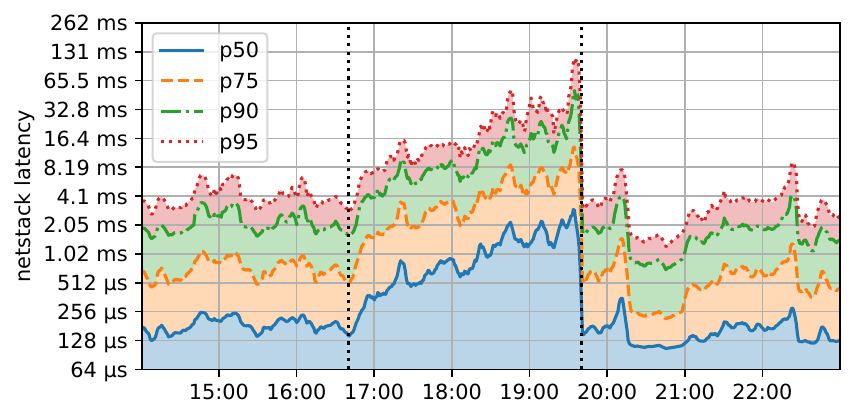}
        \Description{Timeseries-plot showing how the median, 75th, 90th, and 95th percentile tcp-socket-read latency evolves over time. During an anomalous period in the middle, all latency metrics steadily increase, reaching an order of magnitude higher values compared to the start of the period, before suddenly returning to normal. }
    }

    \subcaptionbox{Netstacklat measurements, RX-handler calls per core (left axis), and RPS (right axis)\label{fig:cdn_anomaly_events}}{
        \includegraphics[width=\columnwidth]{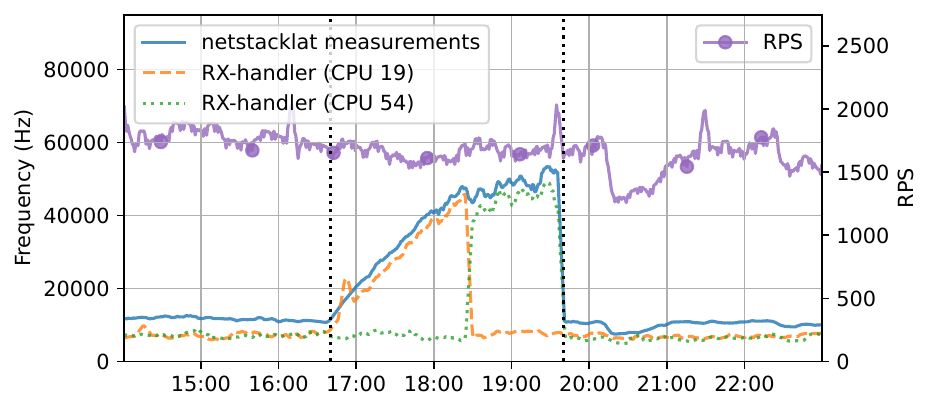}
        \Description{Timeseries-plot showing the frequency of netstacklat measurements and calls to the kernel's packet receive-handler on two cores, as well as the RPS handled by the server. The RPS remains relatively stable throughout the period, while the netstacklat measurements and receive-handler calls show a very similar increase during the anomaly.}
    }

    \caption{Anomalous event where host latency increased drastically before returning to normal. The vertical dotted lines indicate the start and the end of the 3 hour long event.}
    \label{fig:cdn_anomaly}
\end{figure}

We have also observed netstacklat's ability to capture anomalous events, where the tcp-socket-read latency increases even when there is no obvious traffic load that might cause it. Figure~\ref{fig:cdn_anomaly} shows one such instance, where the tcp-socket-read latency increased drastically over a 3-hour period. In Figure~\ref{fig:cdn_anomaly_latency}, we can see how the median tcp-socket-read latency for a server stays within the 128-256us bin before the anomaly starts. At 16:45, the latency then starts increasing steadily, exceeding 2ms towards the end of the anomaly at 19:45, and then quickly falling back to its prior 128-256us range. The tail latency also sees a similar increase, with the 95th percentile increasing from around 2-4ms to briefly exceeding 65ms. 

While the anomaly is clearly visible from the latency increase reported by netstacklat, there is no considerable increase in the request rate during the anomalous period, as shown by the circle-marked line in Figure~\ref{fig:cdn_anomaly_events}. Other traffic metrics, like packet rate and data rate, as well as the overall CPU utilization, also stay similar throughout the anomaly. However, we find that in addition to the tcp-socket-read latency increasing, the number of tcp-socket-read measurements sees a 5-fold increase during the anomaly, shown by the solid line in Figure~\ref{fig:cdn_anomaly_events}. As each measurement corresponds to nginx-ssl reading from a socket, this indicates a drastic increase in socket reads. After further investigation, we find that the number of receive-handler calls in the early part of the kernel network stack running on one specific CPU core (reported by the \texttt{softnet\_data.processed} counter~\cite{linux-dev.c-softnet}), shown by the dashed line, closely mirrors the increasing number of socket reads. Roughly halfway through the anomaly, the high number of receive-handler calls migrates to another CPU core, shown by the dotted line.

The large increase in receive-handler calls on a single core without a corresponding increase in network traffic suggests that Generic Receive Offload (GRO)~\cite{thekerneldevelopmentcommunitySegmentationOffloadsLinux2025} failed to merge SKBs for an elephant flow. GRO is normally supposed to merge multiple smaller SKBs to reduce the overhead in the higher layers of the stack. Without the SKB merging, many small SKBs need to be processed, causing a high network processing load on the core handling that flow. The many small SKBs also cause new data to frequently become readable from the socket, resulting in the high number of socket reads observed by netstacklat. Monitoring the network stack latency was instrumental in detecting this anomaly, as most traditional system metrics did not show any obvious deviations.

We have also observed other anomalies where the tcp-socket-read latency is unexpectedly high. For instance, one server showed consistently higher latency than other servers at the same location, despite not handling any more traffic. An investigation revealed that the server with the high latency was part of an experiment where different CPU isolation mechanisms were evaluated. In this case, the nginx-ssl instance was running on a set of heavily loaded CPU cores, causing the tcp-socket-read latency to increase. There are also several instances where the tcp-socket-read latency during brief periods reaches 10s of milliseconds, where further investigation is required to reveal the underlying cause.

\section{Limitations and future work}
\label{sec:limitations}

While we in this work have shown how the current iteration of netstacklat can efficiently monitor the latency at different layers in the kernel network stack, there are still several limitations that could be addressed in future work to improve netstacklat's usefulness. In particular, netstacklat only captures the latency from the start of the kernel network stack, which does not include the time it takes to transfer the packet from the NIC to the CPU and the delay for the kernel to start processing the packet once transferred. As prior work has shown that congestion in the NIC-CPU interconnect~\cite{agarwalUnderstandingHostInterconnect2022, agarwalHostCongestionControl2023, vuppalapatiUnderstandingHostNetwork2024} and the interrupt-based packet processing in Linux~\cite{liTalesTailHardware2014, gallenmullerDuckedTailsTrimming2021, caiKernelVsUserlevel2023} can cause significant latency, measuring the total host latency from when packets first arrive at the NIC would be highly useful. As many modern NICs can supply hardware timestamps for received packets, netstacklat could be extended to track the full host latency by calculating the delay relative to the hardware timestamp  
(also available as part of the SKB) 
instead of the kernel's software timestamp. 
One complicating factor here is that the NIC hardware timestamps are based on the NIC's own clock, while the eBPF programs use the system clock to measure when an SKB reaches a probe point. To facilitate accurate delay calculations while using timestamps from different clocks, clock conversion factors could be tracked similarly to how NSight~\cite{haeckiHowDiagnoseNanosecond2022} does.

Additionally, netstacklat is solely focused on monitoring the latency in the ingress path. To also monitor latency for egress packets, netstacklat could hook into relevant functions in the kernel's transmission path. 
While the SKB timestamp cannot be used in the same way for transmitted packets, additional eBPF programs could be used to record reference timestamps at entry points of the transmission path, e.g, at \texttt{tcp\_sendmsg()}. However, recording and looking up these separately managed timestamps would add additional overhead compared to the ingress path.  

Netstacklat's ingress monitoring can also be expanded to cover additional transport protocols or different layers of the network stack 
(beyond the ones listed in Section~\ref{sec:netstacklat-hooks})
if needed by identifying suitable kernel functions to probe. 
While we currently make some adjustments for how netstacklat handles the different handpicked probe points to support some ancillary features, such as the TCP HoL filter, the core logic for recording the latency remains the same across all probe points and only requires access to the SKB timestamp. Further generalizing netstacklat's code to allow it to probe any function that receives an SKB, similarly to pwru and retis, could thereby provide more flexible coverage across different parts of the network stack.

For some systems and applications, there can also be network-related processing outside of the kernel network stack that would be relevant to cover. 
% However, with netstacklat relying on both eBPF and the kernel's packet timestamping, it is currently limited to the Linux network stack.
For user space transport and application layer protocols running on top of the kernel network stack, such as QUIC running over UDP, netstacklat will only partially capture the latency. While the delay until the SKB payload is read from the kernel's socket is captured, any additional delays from user space protocol processing before the message is delivered to the final application will not be visible. 
For kernel bypass technologies, such as the Data Plane Development Kit (DPDK)~\cite{dpdk-webpage}, traffic will completely bypass netstacklat's monitoring in the kernel network stack. 
In theory, we could use uprobes~\cite{ebpfdocs-kprobe} or user statically defined tracepoints~\cite{ebpfdocs-usdt} to probe user space stacks without needing to modify any applications. However, such an approach requires that suitable points are traceable and expose a packet timestamp to use as a reference for the latency calculation, greatly limiting its applicability. 

In practice, covering user space components will likely require adding explicit instrumentation in user space libraries, similarly to how Fathom instruments RPC libraries~\cite{qureshiFathomUnderstandingDatacenter2023}. 
Protocols running on top of the kernel network stack must also be modified to read and expose the SKB timestamp, while kernel bypass techniques must provide some alternative form of packet timestamp, such as the \texttt{rte\_dyn\-flag\_rx\_time\-stamp} in DPDK. To facilitate consistent instrumentation in a format compatible with the one used for netstacklat's kernel-based monitoring across different user space protocols and network stacks, it could be beneficial to implement a reusable netstacklat instrumentation library.

While netstacklat in many cases already achieves reasonably low overhead (e.g., less than 0.04\% of the CPU utilization at Cloudflare), capturing measurements multiple times per packet will inherently lead to considerable overhead at very high packet rates. To further reduce the overhead of netstacklat, support for random sampling could be added. While sampling will make it harder to capture the far ends of the latency tails (e.g., the 99.99999999th percentiles shown in Figure~\ref{fig:alltests_ecdf}), even sampling as few as 1:128,000 packets can still provide statistically meaningful results when deployed at large scales~\cite{qureshiFathomUnderstandingDatacenter2023}. However, as a small part of the eBPF program will still need to run each time to make the sampling decision, the overhead will to some extent still scale with the packet rate regardless of the sampling rate.

Finally, we note that while netstacklat can detect the presence of issues causing high network stack latency, its capability to identify the root cause is limited, as it only provides high-level metrics. In general, we would recommend using the high-level overview provided by netstacklat to guide the collection of more detailed traces for root cause identification using complementary tools such as pwru or retis. However, for rare or temporary issues, such as the anomaly in Section~\ref{sec:cdn-anomaly}, it may not be possible to capture the issue outside of the continuous monitoring. Therefore, it could be useful to allow netstacklat to export detailed traces for a small subsample of the packets in addition to capturing the overall latency distribution. Since the design of netstacklat allows the latency to be calculated directly in the eBPF programs, it would be possible to steer the trace generation to packets encountering high latency instead of relying on random sampling to catch the issue.

\section{Conclusion}
\label{sec:conclusion}
In this work, we have presented netstacklat, a novel monitoring tool that collects latency distributions within several layers of the host network stack for ingress traffic. By utilizing eBPF, in-kernel aggregation, and the kernel's existing packet timestamping capabilities, netstacklat achieves compatibility with a wide variety of systems and a low enough overhead to support continuous monitoring in production environments. Through testbed experiments, netstacklat shows how a heavily loaded host can exhibit long latency tails, reaching hundreds of microseconds %even 
in the early parts of the network stack, tens of milliseconds before the packet is enqueued to the socket, and seconds before the application reads the data. Furthermore, we demonstrate that netstacklat has reasonably low overhead, from less than 0.01\% up to 3.2\% CPU utilization across a wide range of HTTP workloads in our testbed experiments, and around 0.04\% CPU utilization during peak hours at Cloudflare's busiest data center location.
% which under heavily loaded scenarios does not inflate end-to-end latency by more than 5\% where other monitoring tools increase it with over 100\%. 
Finally, we share early insights from deploying netstacklat at Cloudflare, among others, highlighting how we through netstacklat found an anomaly where a temporary GRO issue on a single CPU core led to drastically increased host network latency. 

As end hosts increasingly become the bottleneck in networks, we believe that tracking the latency within the host network stack will become a crucial aspect of performance monitoring. While the networking community has already identified several important sources of latency in the current network stacks, we would like to encourage it to also develop more tools to monitor the host latency as network stacks evolve. Here, we hope netstacklat can provide a solution for the Linux network stack, but further work is needed to provide more comprehensive coverage. We plan to continue studying how host network stack latency varies for different workloads and explore its potential as a load indicator.


\begin{acks}

This research was funded by Red Hat Research and the Swedish Knowledge Foundation (grant 20220072). Additional thanks to Felix Maurer for assisting with lattrace, and to the anonymous reviewers for their helpful feedback.

\end{acks}

% \printbibliography
\bibliographystyle{ACM-Reference-Format}
\bibliography{netstacklat}

%\newpage

\appendix

\section{Ethics}
This work does not raise any ethical issues.

\section{Additional details from testbed experiments}
We here share some additional information about the results from the experiments in the testbed covered in Section~\ref{sec:workload-comparison}. This is only intended as supplementary information for the interested reader and is not needed to underpin the key results presented in the paper.

\subsection{Latency distributions}

\begin{figure}
    \centering
    \subcaptionbox{ip-start}{
        \includegraphics[width=\columnwidth]{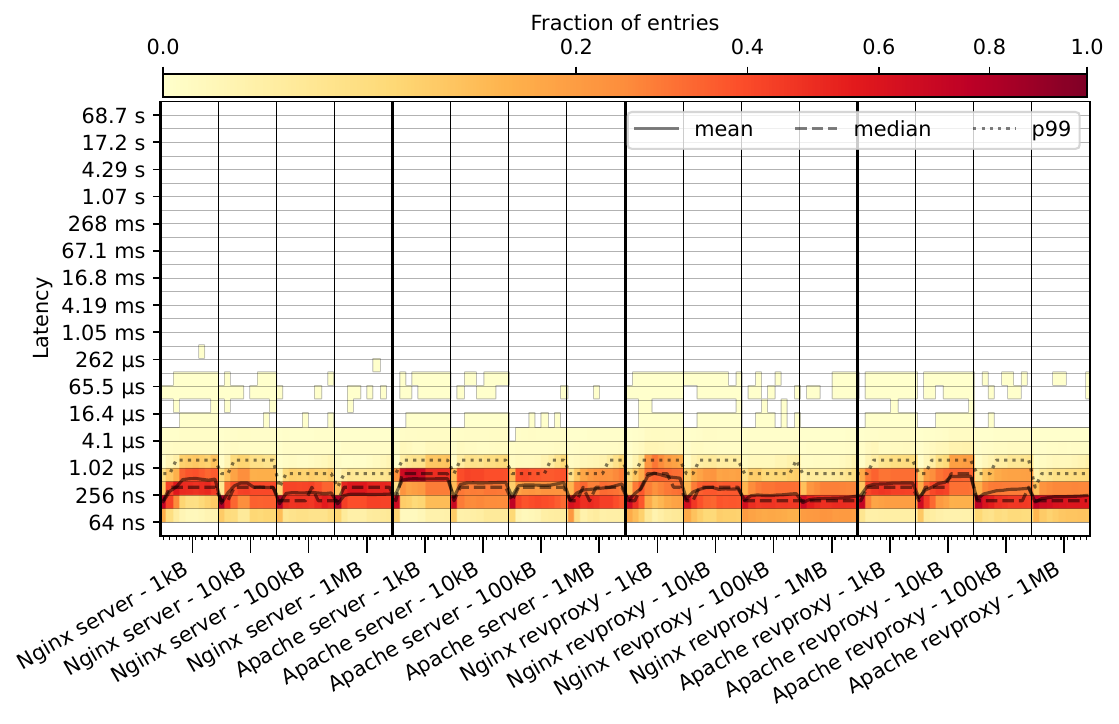}
    }
    \subcaptionbox{tcp-start}{
        \includegraphics[width=\columnwidth]{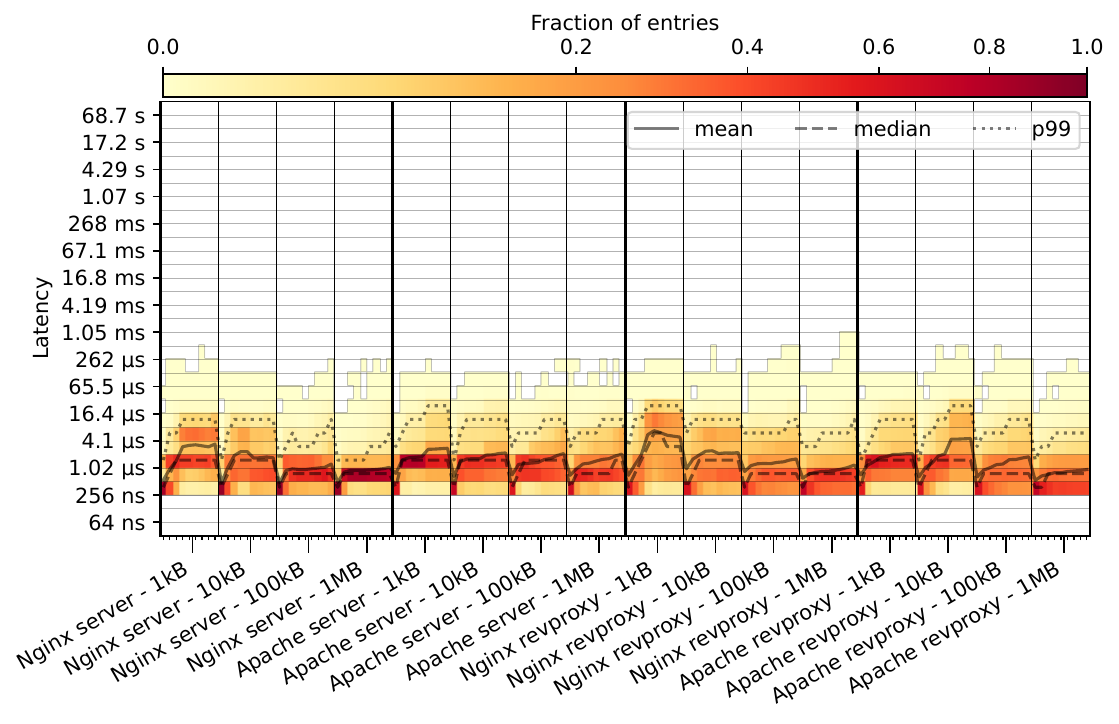}
    }
    \subcaptionbox{tcp-socket-enqueued}{
        \includegraphics[width=\columnwidth]{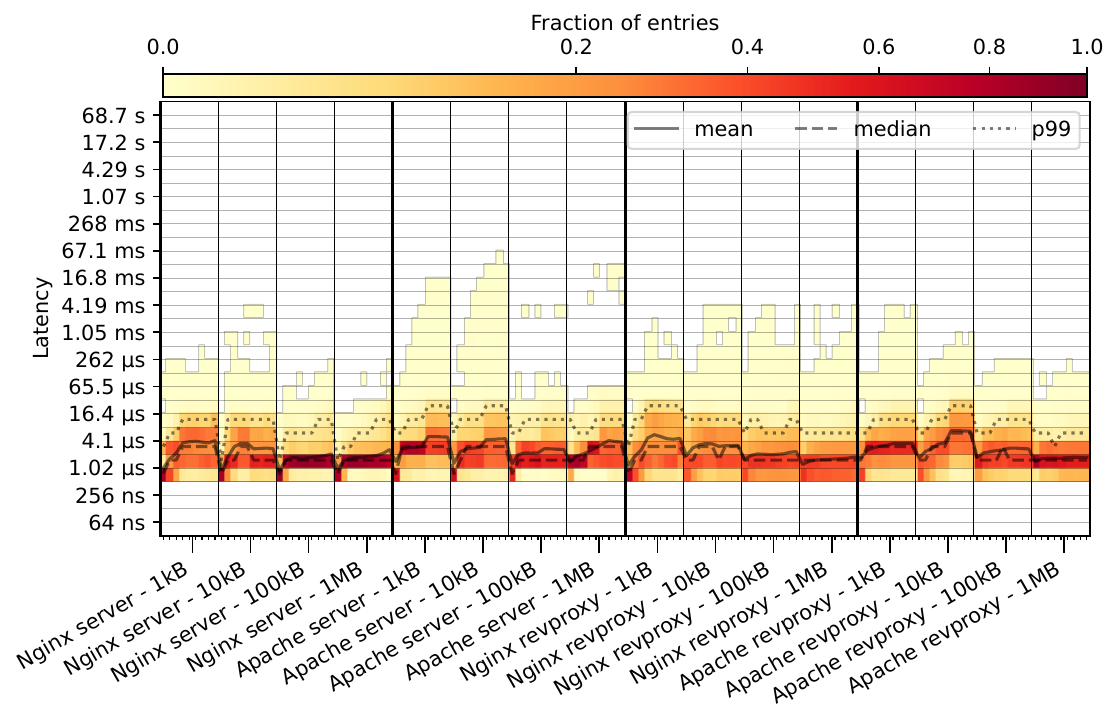}
    }
    \caption{The latency distribution for each of the 144 workloads. Same x-axis layout as Figure~\ref{fig:alltests_latency}.}
    \label{fig:apdix_workloads_latency_dist}
\end{figure}

Figure~\ref{fig:apdix_workloads_latency_dist} shows the full distribution of the latency for ip-start, tcp-start, and tcp-socket-enqueued, which is summarized by the statistics in Figure~\ref{fig:alltests_latency_summary}. The distribution for tcp-socket-read can be found in Figure~\ref{fig:alltests_tcpread_latency}. The full distributions in Figure~\ref{fig:apdix_workloads_latency_dist} show that long latency tails, far exceeding the 99th percentile and consequently not visible in Figure~\ref{fig:alltests_latency_summary}, exist for the earlier hooks across most workload combinations.

\begin{figure}
    \centering
    \subcaptionbox{load generator facing interface}{
        \includegraphics[width=\columnwidth]{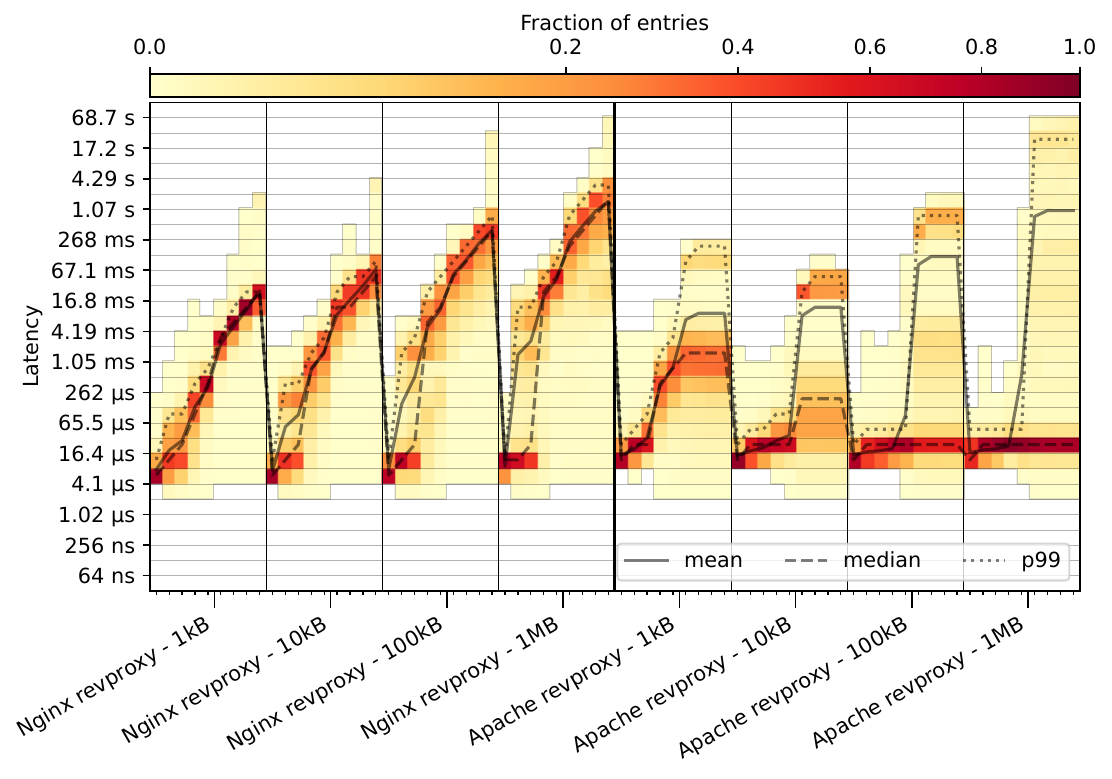}
    }
    \subcaptionbox{origin server facing interface}{
        \includegraphics[width=\columnwidth]{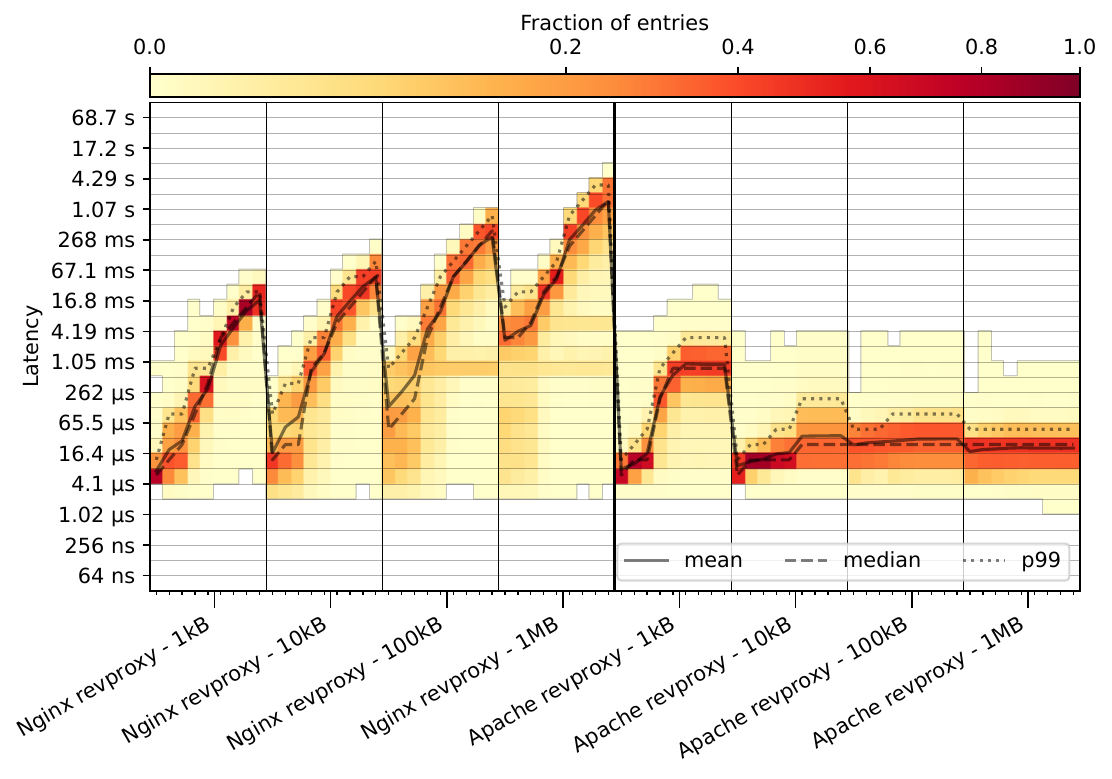}
    }
    \caption{Per-interface tcp-socket-read latency for each of the 72 reverse proxy tests. Same x-axis layout as Figure~\ref{fig:alltests_latency} (although only including the reverse proxy server configurations).}
    \label{fig:apdix_workloads_tcpread_dist_pernic}
\end{figure}

For the experiments using reverse proxy configurations, there will in addition to the incoming requests on the interface connected to the load generator, also be incoming traffic from the origin server containing the requested resources. Figure~\ref{fig:apdix_workloads_tcpread_dist_pernic} shows how the distributions differ between the two sides. For the Nginx reverse proxy case, the latency distribution is similar on both sides. The host is overloaded, so both the traffic from the load generator and the origin server suffer large latencies. 

On the other hand, for the Apache reverse proxy, the latency distributions on each side differ drastically. All the incoming data from the origin server is read swiftly, but some of the requests from the load generator still have to wait a long time. As mentioned in Section~\ref{sec:workload-latency}, for the Apache reverse proxy experiments with files larger than 1kB, the bottleneck is shifted towards the origin server. The DuT therefore has sufficient CPU resources to quickly handle all of the incoming traffic. However, as Apache can only handle a limited number of requests at a time (we use the default limit of 175 connections), some of the requests still have to wait a long time before the Apache reverse proxy can serve them, and Apache has to wait on the origin server to complete its currently handled requests.

\subsection{End-to-end performance}
\begin{figure}
    \centering
    \includegraphics[width=\columnwidth]{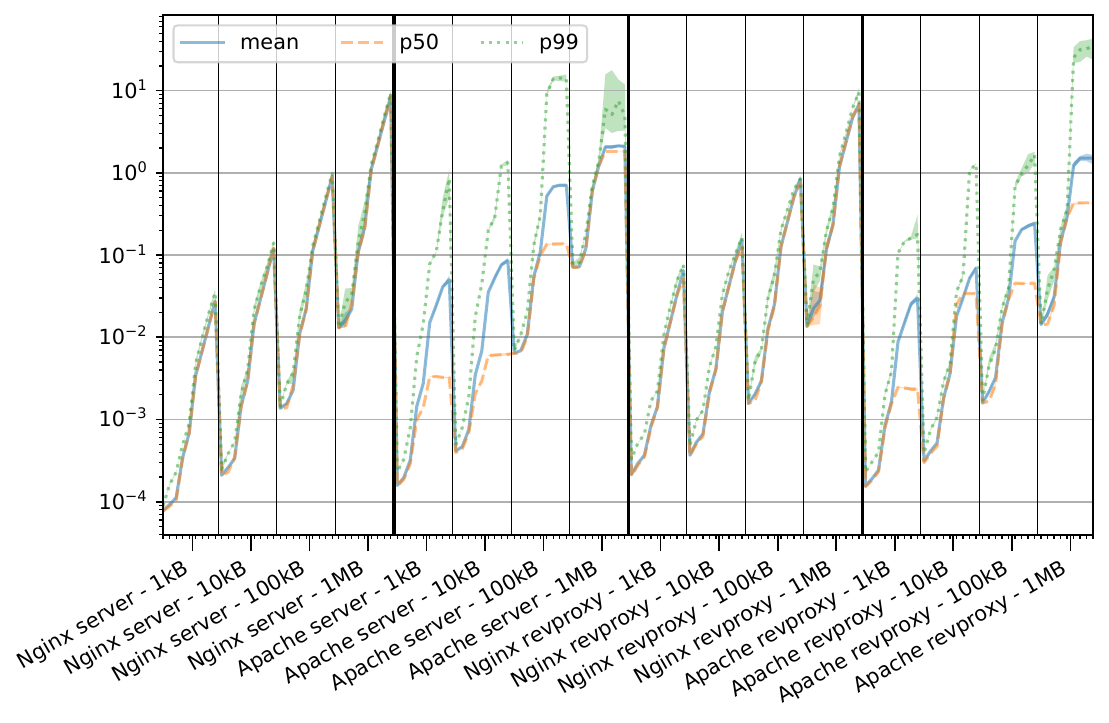}
    \caption{End-to-end latency reported by the load generator for each of the 144 workloads. The line shows the mean across all repetitions, and the shaded area indicates the 95\% confidence interval. Same x-axis layout as Figure~\ref{fig:alltests_latency}.}
    \label{fig:apdix_e2e_latency}
\end{figure}

Figure~\ref{fig:apdix_e2e_latency} shows the end-to-end response time statistics reported by the load generator. We note that the latency distribution here appears to closely match the latency distribution for tcp-socket-read in Figure~\ref{fig:alltests_latency}. That this is the case is also indicated by the strong correlation between the end-to-end latency and the tcp-socket-read latency in Figure~\ref{fig:correlation_matrix}.

The latency for Nginx, either as a server or reverse proxy, has a very narrow distribution where the mean, median, and 99th percentile largely overlap. The latency also keeps increasing as the number of concurrent connections grows all the way up to 4000 connections. Meanwhile, Apache shows a larger variation it its latency distribution, with a clear distinction between the mean, median and 99th percentile. Furthermore, the latency increase for Apache starts to taper of past 1000 concurrent connections as it drops the excess requests it cannot handle.

\begin{figure}
    \centering
    \includegraphics[width=\columnwidth]{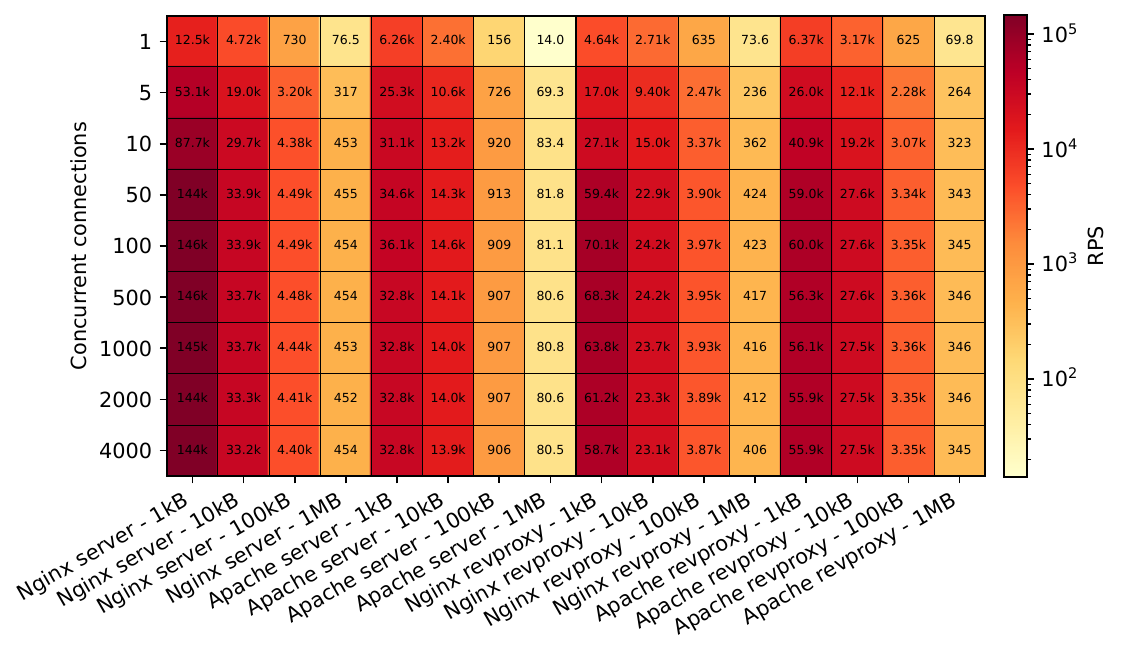}
    \caption{Average request rate reported by the load generator for each of the 144 workloads. SI-prefixes are used to denote the scale in a compact manner.}
    \label{fig:apdix_e2e_rps}
\end{figure}

Figure~\ref{fig:apdix_e2e_rps} shows the average RPS achieved for the different workload combinations. As can be seen, it varies greatly based on both server configuration and file size. This is the reason we opted to vary the load in terms of concurrent connections rather than target request rates. A target request rate of 100 RPS will generate a very light load for Nginx serving 1kB files, but will overload the DuT if we instead request Apache to serve 1MB files. Across all server and file size configurations, we can note that at 50 concurrent connections the system is saturated, and further increasing the number of parallel connections does not increase the achieved RPS.

\subsection{System load}

\begin{figure}
    \centering
    \includegraphics[width=\columnwidth]{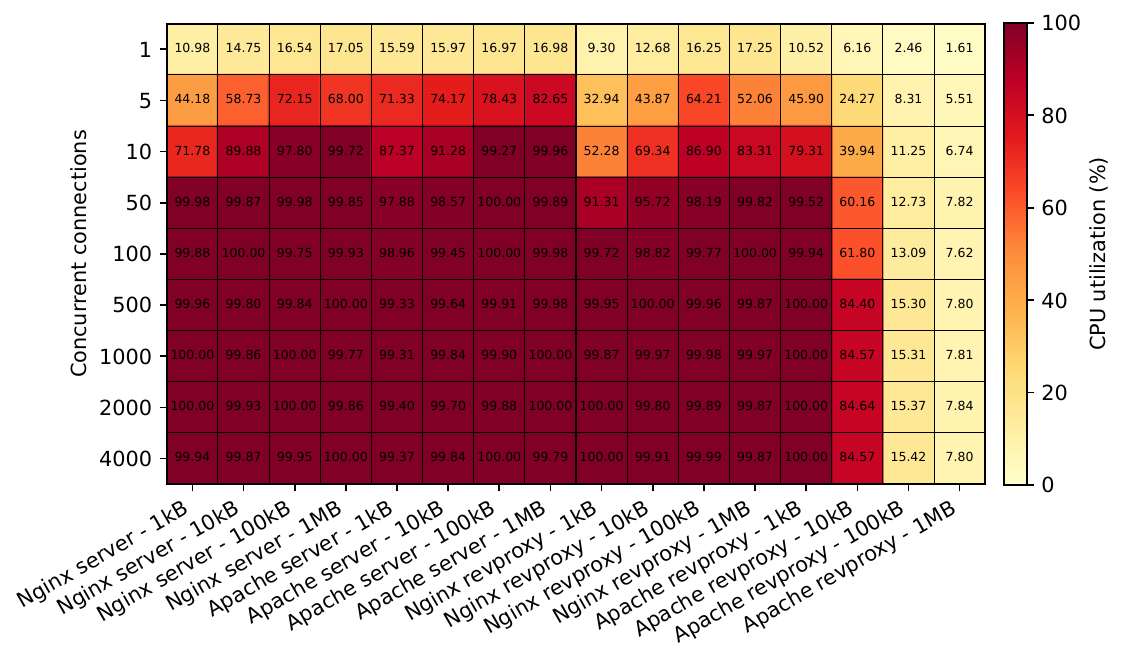}
    \caption{Average CPU utilization on the DuT for each of the 144 workloads.}
    \label{fig:apdix_sys_cpu}
\end{figure}

\begin{figure}
    \centering
    \includegraphics[width=\columnwidth]{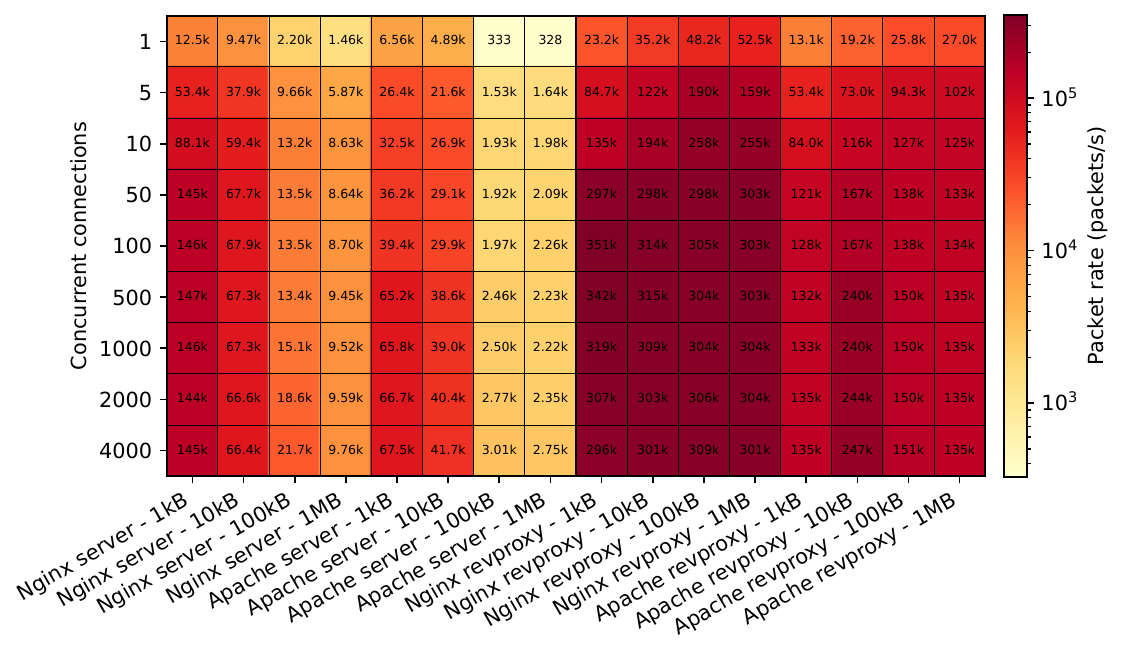}
    \caption{Average received packet rate at the DuT for each of the 144 workloads. SI-prefixes are used to denote the scale in a compact manner.}
    \label{fig:apdix_sys_rx_pps}
\end{figure}

\begin{figure*}
    \centering
    \includegraphics[width=0.9\textwidth]{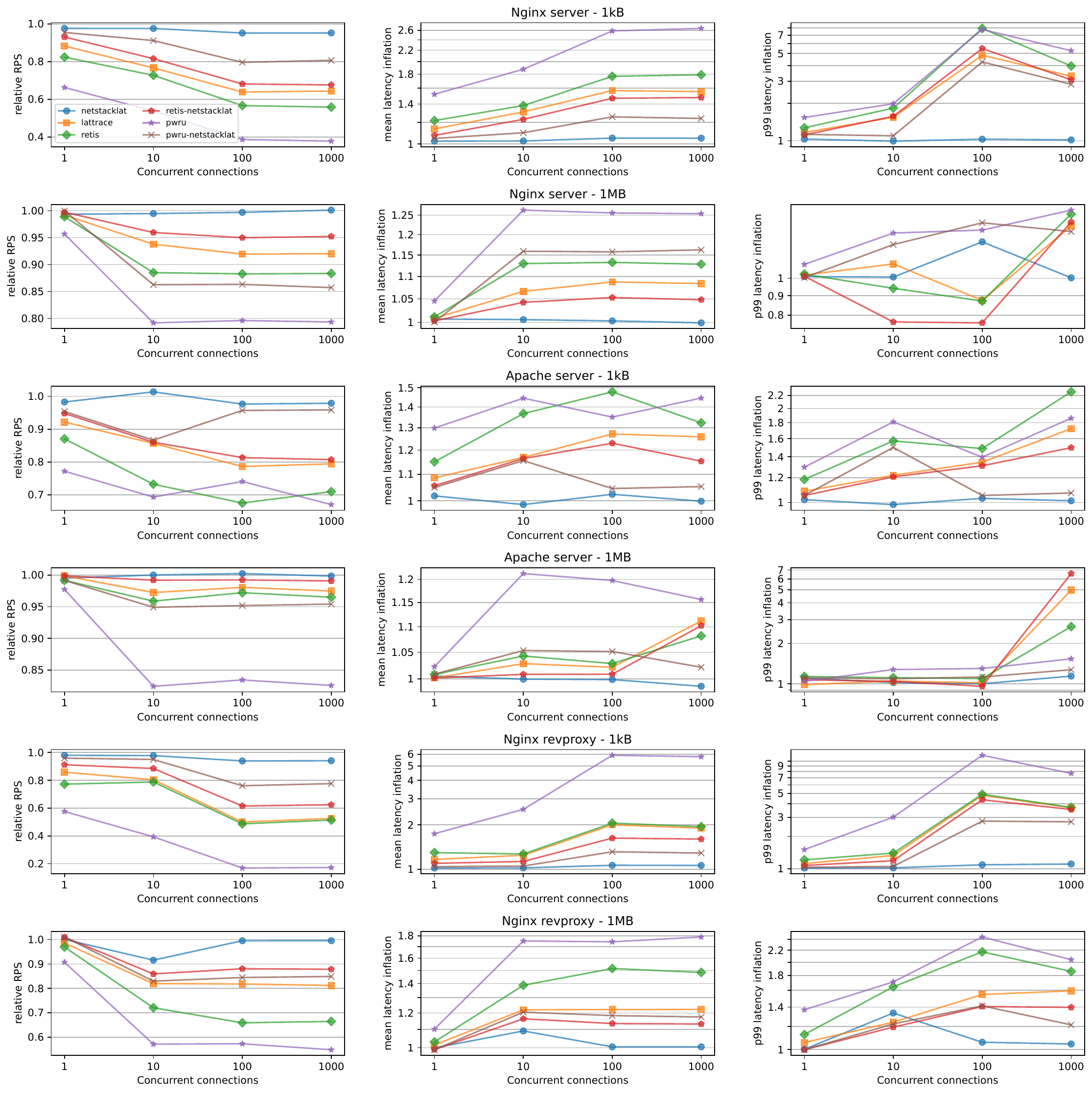}
    \caption{The relative end-to-end performance when running a monitoring tool on the DuT compared the baseline of no monitoring. Each row shows one server configuration and file size combination. The columns, from left to right, show the relative difference in end-to-end RPS, mean response time, and 99th percentile response time, respectively.}
    \label{fig:apdix_tools_e2e_relbaseline}
\end{figure*}

Figure~\ref{fig:apdix_sys_cpu} shows how the average CPU utilization on the DuT varies across the different workloads. Past 50 concurrent connections the DuT is fully saturated in all cases except for the Apache reverse proxy setup serving 10kB or larger files, where the bottleneck is instead shifted towards the origin server. % The reason that the Apache reverse proxy shifts the bottleneck to the origin server while the Nginx reverse proxy is due to their default settings for requesting remote resources. Nginx uses HTTP 1.0 by default, while Apache uses HTTP 1.1.

Figure~\ref{fig:apdix_sys_rx_pps} instead shows the network load in terms of the average ingress packet rate. Here, we can see that the Nginx reverse proxy setup is the configuration that results in the highest network load. As netstacklat's overhead scales roughly proportionally to the packet rate, or more specifically to the rate of processed SKBs, Nginx reverse proxy is also the configuration that has the highest overhead in Figure~\ref{fig:alltests_cpu_perhook}.

\section{Tool comparisons for additional workload configurations}

In Section~\ref{sec:tool-comparison} we focus on a single server configuration and file size combination: Nginx with 10kB files. To verify that the relative performance difference between the evaluated tools is similar for other server configuration and file size combinations, we perform additional experiments with three different sever configurations (Nginx server, Apache server, and Nginx reverse proxy), and two file sizes (1kB and 1MB). Due to the large number of experiments required to test each combination with each tool, we only run a single iteration (instead of 9) for each load test.

Figure~\ref{fig:apdix_tools_e2e_relbaseline} summarizes the impact the evaluated tools have on the end-to-end performance for the six server configuration and file size combinations. Rather than showing bar plots with the total performance, as in Figure~\ref{fig:tools_e2e}, we here normalize the end-to-end performance while running each tool against the baseline performance obtained when no tool runs on the DuT. This provides an overview of the relative performance difference between the tools. Note that with only a single test for each tool and workload combination, we cannot provide any confidence intervals for these results. This becomes especially problematic for the 99th percentile response time when using 1MB files, as they are based on a small number of requests (the slowest 1\% of the 826 - 27437 requests served during the single load test for each combination of server configuration, number of concurrent connections and tool for 1MB files). 
% For Nginx with 1MB files, the small sample size leads to the illogical result that some tools reduce the 99th percentile response time.
The results with 1kB files can be expected to be more stable, as the load tests with 1kB files on average generate over 200 times as many requests as their 1MB counterparts.

Despite the uncertainty in the results, we can note that netstacklat consistently has a very small impact on the end-to-end performance, only deviating from the baseline performance by more than 10\% in a few instances for the 99th percentile latency with 1MB files (where the results are highly uncertain, as explained above). In contrast, all the other tools frequently result in much larger degradations in performance, especially for the load tests with 1kB files, which generate a relatively large amount of ingress traffic. During the heavily loaded scenarios with 100 or 1000 concurrent connections for Nginx as either a server or reverse proxy and serving 1kB files, the other tools inflate 99th percentile latency by at least 170\%. Across the $6 * 4 = 24$ combinations of workloads and number of concurrent connections shown here, netstacklat on average reduces RPS by 2\% and inflates the mean and 99th percentile latency with 2\% and 6\%, respectively. In comparison, pwru-netstacklat, which has the second lowest impact, on average reduces RPS by 10\%, inflates mean latency by 12\% and inflates 99th percentile latency by 50\%. The main finding from Section~\ref{sec:tool-e2e-impact}, that netstacklat has a much smaller impact on the end-to-end performance compared to the other tools, thus appears to generalize across different workload and server configurations.

\end{document}